\documentclass[12pt]{article}
\usepackage{amsfonts,latexsym,amscd}

\usepackage{graphics}
\usepackage[usenames]{color}

\newtheorem{theorem}{Theorem}[section]

\newtheorem{proposition}[theorem]{Proposition}
\newtheorem{corollary}[theorem]{Corollary}
\newtheorem{lemma}[theorem]{Lemma}

\newcommand{\calG}{{\mathcal{G}}}
\newcommand{\calD}{{\mathcal{D}}}
\newcommand{\calB}{{\mathcal{B}}}

\newcommand{\dis}{\displaystyle}
\newcommand{\rarrow}{\rightarrow}
\newcommand{\cd}{\:\cdot\:}
\renewcommand{\theequation}{\thesection.\arabic{equation}}

\def\boldfacefake #1{%
       \hbox{%
              \mathsurround=0pt 
              \hbox to 0.5pt{$#1$\hss}%
              \hbox to 0.5pt{$#1$\hss}%
              \hbox {$#1$}%
       }%
}
\def\boldone{{\boldfacefake{1}}}

\begin{document}

\title{End-to-end Distance from the Green's Function for a Hierarchical Self-Avoiding Walk in
Four Dimensions}
\author{
David C. Brydges\thanks{Research supported by NSF Grant DMS-9706166}\\
  	University of British Columbia\\
	Mathematics Department\\
	\#121-1984 Mathematics Road \\ 
	Vancouver, B.C. V6T 1Z2\\  
	Canada\\
{\tt db5d@math.ubc.ca}\\[1mm] 
and\\[1mm]
Department of Mathematics \\ 
Kerchof Hall \\ 
P. O. Box 400137 \\ 
University of Virginia\\[3mm] 
\and
John Z. Imbrie \\
Department of Mathematics \\ 
Kerchof Hall \\ 
P. O. Box 400137 \\ 
University of Virginia\\ 
Charlottesville, VA 22904-4137\\
{\tt ji2k@virginia.edu}
}

\date{}
\maketitle
\thispagestyle{empty}

\begin{abstract}
In \cite{BEI} we introduced a Levy process on a hierarchical lattice which is four dimensional, in the sense that the Green's function for the process equals $\frac{1}{|x|^2}$. If the process is modified so as to be weakly self-repelling, it was shown that at the critical killing rate (mass-squared) $\beta^c$, the Green's function behaves like the free one.

Now we analyze the end-to-end distance of the model and show that its expected value grows as a constant times $\sqrt{T}\log^{\frac{1}{8}}T \left(1+O\left(\frac{\log\log T}{\log T}\right) \right)$, which is the same law as has been conjectured for self-avoiding walks on the simple cubic lattice $\mathbb{Z}^4$. The proof uses inverse Laplace transforms to obtain the end-to-end distance from the Green's function, and requires detailed properties of the Green's function throughout a sector of the complex $\beta$ plane. These estimates are derived in a companion paper \cite{BI2}.
\end{abstract}

\newpage
\pagenumbering{roman}

\tableofcontents

\newpage

\pagenumbering{arabic}
\setcounter{page}{1}

\section{Introduction}

\subsection{Main results}
\renewcommand{\theequation}{\thesection.\arabic{equation}}

Precise calculations by theoretical physicists have established, with the aid of some reasonable assumptions, that the end-to-end distance of a self-avoiding walk at time $T$ should be asymptotic to a constant times $T^{\frac{1}{2}} \log^{\frac{1}{8}}T$ as $T$ tends to infinity.  See for example \cite{BLZ} and additional references in \cite{MS}. These arguments form a starting point for complete proofs. In our previous paper \cite{BEI} of this series, we started such a program but with two major simplifications. The first is to study processes which repel weakly as opposed to being strictly self-avoiding. The second is to replace the simple cubic lattice by another state space, a ``hierarchical lattice,'' specifically designed to facilitate the use of the renormalization group. While the renormalization group is proposed for proving these results also on the simple cubic lattice, the method is considerably simpler to apply on the hierarchical lattice.

The hierarchical lattice and some of its history have been described at length in \cite{BEI}. Here we summarize that discussion and specialize it to four dimensions. The hierarchical lattice $\calG$ is the direct sum of infinitely many copies of $\mathbb{Z}_n$, where $n = L^4$ for some integer $L >1$ which characterizes the lattice. A typical element $x \in \calG$ has the form $x = (\ldots, x_2,x_1,x_0)$ with $x_i \in \mathbb{Z}_n = \{0,1,\ldots,n-1\}$. All but finitely many elements of the sequence $x$ vanish. Let $x_{N-1}$ be the first element, reading from the left, which does not vanish. We define a $\calG$-invariant ultra-metric on $\calG$ by
\begin{equation}\label{equation1.1}
\mbox{dist}(x,y) \equiv |x-y|, \quad |x| \equiv \left\{\begin{array}{ll}
0 & \mbox{if } x = (\ldots 0) \\[2mm]
L^N & \mbox{if } x = (\ldots,x_{N-1},x_{N-2},\ldots,x_0).
\end{array}\right.
\end{equation}

Let $\omega(t)$ be a Levy process on $\calG$ such that
\begin{equation}\label{equation1.2}
P(\omega(t+dt) = y|\omega(t) = x) = C|x-y|^{-6}dt,
\end{equation}
if $x \neq y$.  In \cite{BEI}, Proposition 2.3, we show that, with the right choice of $C = C(L)$, the 0-potential (Green's function) for this process is given by
\begin{eqnarray}\label{equation1.3}
G_0(x-y) & \equiv &
\int^\infty_0 dT \, E_x(\boldone_{\{ \omega(T)=y \} }) \nonumber \\[4mm]
& = & \left\{\begin{array}{ll}
\dis\frac{1-L^{-4}}{1-L^{-2}} & \mbox{ if } x = y; \\[3mm]
\dis\frac{1}{|x-y|^2} & \mbox{ if } x \neq y.
\end{array}\right.
\end{eqnarray}
The process $\omega(t)$ is ``four dimensional'' in the sense that its Green's function is $\frac{1}{|x-y|^2}$ for $x \neq y$. The slow decay in the law (\ref{equation1.2}) is an ugly contrast with the simplicity of the nearest neighbor random walk on the simple cubic lattice, but it is a necessary price for a state space with an ultra-metric. (On such a space, a process with finite range jumps cannot leave the ball whose radius equals the range and which is centered on the starting position.) One consequence of (\ref{equation1.2}) is that $\omega(t)$ does not have second moments. Thus we will measure end-to-end distance by $E_0(|w(T)|^\alpha)^{\frac{1}{\alpha}}$ with $0 < \alpha < 2$. At first one might expect 
that if this quantity is normalized by $\frac{1}{\sqrt{T}}$ it would have a limit as $T \rarrow \infty$.
Instead the behavior is asymptotically periodic in log $T$, as the following proposition shows.

\begin{proposition}\label{proposition1.1}
Fix $L > 1$. Then for each $\alpha$, $0 < \alpha < 2$, and each $T \geq 0$,
$$
\mathop{\rm lim}\limits_{m \rarrow \infty} \frac{1}{\sqrt{L^{2m}T}} E_0(|\omega(L^{2m}T)|^\alpha)^{\frac{1}{\alpha}}
$$
exists and is a strictly positive, {\em non-constant}, bounded function $F_\alpha(T)$ which satisfies
$F_\alpha(L^2T) = F_\alpha(T)$.
\end{proposition}

We postpone the proof of this proposition and turn our attention to the self-repelling process. Let
us define $\tau(x) \equiv \tau^{(T)}(x)$ as the local time (up to time $T$) that the process spends at state $x$: 
\begin{equation}\label{equation1.4}
\tau^{(T)}(x) \equiv \int^T_0 ds \, \boldone_{\{\omega(s) =x\}}.
\end{equation}
Let
\begin{equation}\label{equation1.5}
\tau^{2}(\calG) \equiv \int_{\calG} dx \, \tau^2(x) = \int ds \, dt \,
\boldone_{\{\omega(s) = \omega(t)\}},
\end{equation}
where $\int dx$ is Haar measure, i.e., counting measure on $\calG$. Clearly, $\tau^2(\calG)$ is a measure of how much time the process spends in self-intersecting. For each choice of a parameter $\lambda \geq 0$ we define a new ``self repelling'' process $\omega_\lambda$ whose expectation $E_{x,\lambda}^T$ is given by
\begin{equation}\label{equation1.6}
E_{x,\lambda}^T(\cd) \equiv \frac{E_x\left(e^{-\lambda \tau^{2}(\calG)}(\cd)\right)}
{E_x\left(e^{-\lambda \tau^{2}(\calG)}\right)}.
\end{equation}
(Recall $\tau = \tau^{(T)}$.) We are able to control this expectation for $\lambda$ in a sector of the complex plane containing the positive reals, although the measure may no longer be real.

The main result of this paper is

\begin{theorem}\label{theorem1.2}
Fix an integer $L \geq 2$ and choose any $0 < \alpha < 2$. If $\lambda$ is sufficiently small
with $|\mbox{arg } \lambda| < \frac{\pi}{3}$, then
\begin{equation}\label{equation1.7}
E^T_{0,\lambda}(|\omega (T)|^\alpha)^{\frac{1}{\alpha}} = \left(1+
\frac{O(\lambda)}{\ell(T^{-1})}\right) E_0 \left(\left|\omega \left(T\ell(T^{-1})^{\frac{1}{4}}\right)\right|^\alpha\right)^{\frac{1}{\alpha}},
\end{equation}
where with $T > 1$, $B \equiv 1-L^{-4}$, the logarithmic factor is
\begin{equation}\label{equation1.8}
\ell(T^{-1}) = 1+O(\lambda)+B\lambda(4 \log T + \log|1+\lambda \log T|).
\end{equation}
\end{theorem}

{\bf Conventions.} In this paper log refers to the base $L$ logarithm. While we can take any $L \geq 2$ as in \cite{BEI}, for simplicity we restrict to the case where $L$ is a fixed, large integer, and $\lambda$ is taken to be sufficiently small, depending on $L$. Proposition II.6.1, in particular, is easier to state under these assumptions.

Theorem \ref{theorem1.2} describes how if a weak repulsion is switched on, the effect relative to the process without repulsion is to rescale time by the slowly varying $\ell(T^{-1})^{\frac{1}{4}}$. Thus if we say that Proposition \ref{proposition1.1} gives a sense in which
\begin{equation}\label{equation1.9}
|\omega(T)| \simeq c\sqrt{T},
\end{equation}
then in an equivalent sense, for some $c(L,\lambda)$,
\begin{equation}\label{equation1.10}
|\omega_\lambda(T)| \simeq c(L,\lambda) \sqrt{T} \log^{\frac{1}{8}}T \left(
1+\frac{\log \log T}{32\log T}+O\left(\frac{1}{\lambda \log T}\right) \right)
\end{equation}
as $T \rarrow \infty$.

\subsection{Green's functions and the end-to-end distance}

We will be using the field-theoretic representation of the self-avoiding walk, see \cite{BEI}. In this representation, the length of the walk $T$ is integrated over, as in (1.3). We may define the Green's function as a Laplace transform as follows:
\begin{equation}\label{equation1.11}
G_\lambda(\beta,x) \equiv \int^\infty_0 dT \, e^{-\beta T} E_0 \left(e^{-\lambda \tau^{2}(\calG)} \boldone_{\{\omega(T)=x\}}\right).
\end{equation}
Then, after obtaining detailed estimates of the behavior of $G_\lambda(\beta,x)$ we can prove Theorem \ref{theorem1.2} by inverting the Laplace transform to recover fixed-$T$ quantities. This is done in Section 3.

To see how this works, consider simple random walk on $\mathbb{Z}^d$, the process whose generator is the lattice Laplacian $\Delta$. For this model we have
$$
G(\beta,x) = \int^\infty_0 dT \, e^{-\beta T} e^{T\Delta}(0,x) = (-\Delta +\beta)^{-1}(0,x).
$$
We may compute
\begin{eqnarray*}
\sum_x G(\beta,x) & = & \frac{1}{p^2+\beta}\bigg|_{p=0} = \beta^{-1}, \\[4mm]
\sum_x x^2G(\beta,x) & = & \sum^d_{j=1} \frac{d^2}{dp^2_j} \ \frac{1}{p^2+\beta}\bigg|_{p=0} = 2d \beta^{-2}
\end{eqnarray*}
(the lattice expressions reduce to these at $p=0$).  Then we may use inverse Laplace transforms to recover the fixed-$T$ quantities. With $a > 0$ we find
\begin{eqnarray*}
\sum_x P(T,x) & = & \int^{a+i\infty}_{a-i\infty}  \frac{d\beta}{2\pi i} e^{\beta T} \beta^{-1}
= 1, \\[4mm]
\sum_x x^2 P(T,x) & = & \int^{a+i\infty}_{a-i\infty}  \frac{d\beta}{2\pi i} e^{\beta T} 2d \beta^{-2} = 2dT.
\end{eqnarray*}
Here we use the residue theorem to evaluate these contour integrals. Now taking the ratio we see that the expected value of $\omega(T)^2$ is $2dT$.

Returning to the model on the hierarchical lattice, note that in \cite{BEI}, {\em c.f.}~p.~85, we studied
\begin{equation}\label{equation1.12}
U_\lambda (a,x) \equiv \lim_{\Lambda \nearrow \calG} \int^\infty_0 dT \, E_0\left(e^{-\lambda \tau^{2}(\Lambda)-a\tau(\Lambda)}\boldone_{\{\omega(T)=x\}}\right),
\end{equation}
where
\begin{eqnarray}\label{equation1.13}
\tau^2(\Lambda) & \equiv & \int_\Lambda dx \, \tau^2(x) = \int^T_0 \int^T_0 ds \, dt \, \boldone_{\{\omega(s)=\omega(t) \in \Lambda\}}, \\[4mm]
\tau(\Lambda) & = & \int_\Lambda dx \, \tau(x) = \int^T_0 ds \, \boldone_{\{\omega(s) \in \Lambda\}}.
\end{eqnarray}
Hence the difference between $U_\lambda$ and $G_\lambda$ lies in whether $\mathop{\rm lim}\limits_{\Lambda \nearrow \calG}$ lies inside or outside $\int dT \, E_0$. In \cite{BEI} it was shown that there exists, for $\lambda$ small, a special value $a_c(\lambda)$ with the property that
$$
U_\lambda (a_c(\lambda),x) \approx \frac{\rm Const.}{|x|^2}, \quad x \rarrow \infty.
$$
Note that at $\lambda = 0$, $a_c(\lambda) = 0$ by (1.3). It is a by-product of this paper that this $a_c(\lambda)$ is the same as $\beta^c(\lambda)$ which appears in the next proposition and that $G_\lambda = U_\lambda$ for $\beta$ in a sector to the right of $\beta^c(\lambda)$.

We study the interacting Green's function $G_\lambda(\beta,x)$ for $(\lambda,\beta)$ in certain complex domains. Let us introduce the notation
\begin{eqnarray}\label{equation1.15}
\calD_\beta & = & \{\beta \neq 0: \ |\mbox{arg } \beta| < b_\beta\}; \nonumber \\[4mm]
\calD_\lambda & = & \{\lambda: \ 0 < |\lambda| < \delta \mbox{ and } |\mbox{arg } \lambda| < b_\lambda\}; \nonumber \\[4mm]
\overline{\calD}_\beta & = & \{\beta \neq 0: \ |\mbox{arg } \beta| < b_\beta
+ \frac{1}{4} b_\lambda +\epsilon\}; \nonumber \\[4mm]
\overline{\calD}_\lambda & = & \{\lambda: \ 0 < |\lambda| < \overline{\delta} \mbox{ and } |\mbox{arg } \lambda| < b_\lambda+\epsilon\}; \nonumber \\[4mm]
\calB(\rho) & = & \{\beta: \ |\beta| < \rho\}; \nonumber \\[4mm]
\overline{\calD}_\beta(\rho) & = & \overline{\calD}_\beta + \calB(\rho),
\end{eqnarray}
where $b_\beta > 0$, $b_\lambda > 0$ are fixed so as to satisfy
$2b_\beta + \frac{3}{2} b_\lambda < \frac{3\pi}{2}$. In particular,
this means $b_\lambda < \pi$, $b_\beta < \frac{3\pi}{4}$. The number
$\epsilon$ is fixed and small enough so that
$2(b_\beta+\epsilon)+\frac{3}{2}(b_\lambda+\epsilon) < \frac{3\pi}{2}$
also. The number $\overline{\delta}$ is chosen to satisfy the
hypotheses of Proposition II.6.1 below, and $\delta < \bar{\delta}$ is
chosen after (depending on $b_\lambda$). { $\rho = \frac{1}{2}$ by default.}
In order to invert the Laplace transform with good bounds we shall require $b_\beta > \frac{\pi}{2}$, and so $b_\lambda < \frac{\pi}{3}$. For example, $(b_\beta,b_\lambda) = \left(\frac{5\pi}{8},\frac{\pi}{8}\right)$ defines an acceptable pair of domains $(\calD_\beta,\calD_\lambda)$. As $b_\beta$, $b_\lambda$, $\epsilon$, $L$ are taken as fixed, we will usually not make explicit the dependence of constants on 
these parameters.
\bigskip

{\bf Remark.} A somewhat larger domain for $(\beta,\lambda)$ defined by the conditions $|2 \mbox{ arg } \beta - \frac{3}{2} \mbox{ arg } \lambda | < \frac{3\pi}{2}$, $|\mbox{arg } \lambda| < \pi$, $|\mbox{arg } \beta| < \pi$ could be used but for simplicity we have taken domains which are in product form.
\bigskip

Our main theorem for $G_\lambda$ refers to a sequence $(\beta_j,\lambda_j)_{j=0,1,\ldots}$ generated by a recursion defined in paper II \cite{BI2}. The following proposition (proven in paper II) gives all the properties of the recursion that will be needed in this paper.

\bigskip
\noindent
{\bf Proposition II.6.1} {\em
Let ${ (\beta_{0},\lambda_{0}) =  } (\beta,\lambda)$ be in the domain $\overline{\calD}_\beta\left(\frac{1}{2}\right) \times \overline{\calD}_\lambda$ with
$\overline{\delta}$ sufficiently small. The sequence 
$(\beta_j,\lambda_j)_{j=0,1,\ldots,M}$ is such that
\begin{eqnarray}\label{equation1.16}
\lambda_{j+1} & = & \lambda_j - \, \frac{8B \lambda^2_j}{(1+\beta_j)^2} + \epsilon_{\lambda,j}, \nonumber \\[4mm]
\beta_{j+1} & = & L^2\left[\beta_j + \frac{2B}{1+\beta_j} \, \lambda_j\right] + \epsilon_{\beta,j},
\end{eqnarray}
where $\epsilon_{\lambda,j}$, $\epsilon_{\beta,j}$ are analytic functions of ${ (\beta,\lambda) }$ satisfying
\begin{eqnarray}\label{equation1.17}
|\epsilon_{\lambda,j}| & \leq & c_L|\lambda_j|^3|1+\beta_j|^{-1}, \nonumber \\[4mm]
|\epsilon_{\beta,j}| & \leq & c_L|\lambda_j|^2|1+\beta_j|^{-2}.
\end{eqnarray}
Here $B = 1-L^{-4}$, and $M$ is the first integer such that $(\beta_M,\lambda_M)$ is not in the domain $\overline{\calD}_\beta\left(\frac{1}{2}\right) \times \overline{\calD}_\lambda$. If no such integer exists, then $M = \infty$.
}
\bigskip

The next proposition constructs the ``stable manifold'' $\beta^c(\lambda)$ for the recursion above.

\begin{proposition}\label{proposition1.3}
For each $\lambda \in \calD_\lambda$ there exists $\beta^c(\lambda) = O(\lambda)$ with the property that $\beta^c_n \equiv \beta_n(\beta^c(\lambda)) = O(\lambda_n(\beta^c(\lambda))) \rightarrow 0$ as $n \rightarrow \infty$.  Furthermore, if $\beta \in \calD_\beta + \beta^c(\lambda)$, then $\beta_n \in \overline{\calD}_\beta + \beta^c_n$ and $\lambda_n \in \overline{\calD}_\lambda$ for all $n$.
\end{proposition}
This $\beta^c(\lambda)$ is called the {\em critical killing rate}. (It is negative if $\lambda >0$.) We define new variables $\hat{\beta} = \beta - \beta^c(\lambda)$ and $\hat{\beta}_j = \beta_j - \beta^c_j$.

We will relate the interacting Green's function $G_\lambda(\beta,x)$, $\lambda \neq 0$, to the free Green's function $G_0(\hat{\beta},x)$. As we shall see, $G_0(\beta,x)$ is analytic in $\beta$ except for a sequence of poles which lie in the interval $[-1,0)$ and which accumulate at zero. For small $|x|$, that is, $|\beta| \, |x|^2 < 1$, it resembles $|x|^{-2}$. For large $|x|$, that is, $|\beta| \, |x|^2 \geq 1$, it decays as $|x|^{-6}$. Thus $G_0$ has ``range'' $\beta^{-\frac{1}{2}}$. Our next result gives the detailed behavior of $G_0$ (see Section 2 for the proof).

\begin{proposition}\label{proposition1.4} The following statements hold for all $\beta \in \overline{\calD}_\beta$.
\begin{itemize}
\item[(1)]
$$
G_0(\beta,x) = \sum_{j \geq 0} L^{-2j} \frac{(1-L^{-4})(1-L^{-2-2j})}
{|x|^2(1+\beta|x|^2L^{-2})(1+\beta|x|^2L^{2j})},\mbox{  } x \neq 0.
$$
\item[(2)]
$$
G_0(\beta,0) = \sum_{j \geq 0} L^{-2j} \frac{1-L^{-4}}{1+L^{2j}\beta}.
$$

\item[(3)]
There are positive ($L$-dependent) constants $c_1$, $c_2$ such that
$$
\frac{c_1}{|x|^2(1+|\beta| \, |x|^2)^2} \leq |G_0(\beta,x)| \leq 
\frac{c_2}{|x|^2(1+|\beta| \, |x|^2)^2},\mbox{  } x \neq 0,
$$

$$
\frac{c_1}{1+|\beta|} \leq |G_0(\beta,0)| \leq 
\frac{c_2}{1+|\beta|}.
$$
\end{itemize}
\end{proposition}

The next theorem shows how well $G_\lambda$ may be approximated by $G_0$. Provided an effective $\beta$ is used for $G_0$, the error in the approximation is proportional to an effective $\lambda$. The proof is based on the renormalization group and the field theory representation for $G_\lambda$. It will be treated in paper II.

\bigskip
\noindent
{\bf Theorem II.1.1} {\em
Let $\lambda \in \calD_\lambda$ with $\delta$ sufficiently small.  Then $G_{\lambda}(\beta,x)$ is analytic in $\beta$ in the domain $\calD_\beta
+\beta^c(\lambda)$ and
\begin{equation}\label{equation1.18}
|G_{\lambda}(\beta,x)-G_0(\beta_{{\rm eff},N(x)},x)| 
\leq O(\lambda_{N(x)})|G_0(\beta_{{\rm eff},N(x)},x)| .
\end{equation}
Here $N(x) = \log |x|$ for $ x \neq 0$, $N(0) = 0$, and $\beta_{{\rm
eff},j} = L^{-2j} \hat{\beta}_j .$
}
\bigskip

As the behavior of of $G_0$ is described accurately in Proposition \ref{proposition1.4}, this theorem gives a correspondingly accurate picture of $G_\lambda$.  We may interpret $\beta_{{\rm eff},N(x)}$ as the value of $\hat{\beta}$ which would evolve to $\hat{\beta}_{N(x)}$ after $N(x)$ steps of the trivial $(\lambda=0)$ recursion $\hat{\beta}_{j+1} = L^2 \hat{\beta}_j$. The integer $N(x)$ is the number of steps needed to ``bring 0 and $x$ together'' when scaling and decimating the lattice as in \cite[p.~99]{BEI}.

The next proposition shows that $\lambda_{N(x)}$ is something like $N(x)^{-1} = (\log |x|)^{-1}$ for $|x| \leq \hat{\beta}^{-\frac{1}{2}}$. Hence the difference $G_\lambda -G_0$ in (\ref{equation1.18}) decays more rapidly than either term by itself (at least out to the range $\approx \hat{\beta}^{-\frac{1}{2}}$).

\begin{proposition}\label{proposition1.5}
For all $(\hat{\beta},\lambda) \in \calD_\beta \times \calD_\lambda$, the following statements hold for $k = 0,1,2,\ldots$:
\begin{itemize}
\item[(1)]
$(\hat{\beta}_k,\lambda_k) \in \overline{\calD}_\beta \times \overline{\calD}_\lambda$.
\item[(2)]
Let $k_{\hat{\beta}}$ be the largest $k$ such that $|\hat{\beta}_k| \leq 1$ (if no such integer exists, then $k_{\hat{\beta}}=0$). Then
$$
k_{\hat{\beta}} = O(1) + \frac{1}{2} \log(1+|\hat{\beta}|^{-1}) + \frac{1}{8} \log |1+4B \lambda \log(1+|\hat{\beta}|^{-1})|.
$$
\item[(3)]
Let $\hat{k} = \min\{k,k_{\hat{\beta}}\}$. Then
$$
\left|\lambda_k - \, \frac{\lambda}{1+8B \lambda \hat{k}}\right| \leq c_1 
\left|\frac{\lambda}{1+8B \lambda \hat{k}}\right|^2(1+\ln(1+|\lambda| \hat{k})).
$$
\item[(4)]
Let $\ell_k(\hat{\beta})^{-\frac{1}{4}} \equiv  \beta_{{\rm eff},k}/\hat{\beta}
= \beta_k L^{-2k}/\hat{\beta}$. Then
$$
\ell_k(\hat{\beta}) = (1+8B \lambda \hat{k})e^{O(\lambda)}.
$$
\item[(5)]
$\lambda_{\hat{\beta}} \equiv \dis\lim_{k \rarrow \infty} \lambda_k$ and $\beta_{{\rm eff},\infty}
\equiv \dis\lim_{k\rarrow \infty} \beta_{{\rm eff},k}$ exist, as does $\ell(\hat{\beta}) \equiv \dis\lim_{k\rarrow \infty} \ell_k(\hat{\beta}) = \beta_{{\rm eff},\infty}/\hat{\beta} = (1+8B \lambda k_{\hat{\beta}})e^{O(\lambda)}$.
\item[(6)]
Let $|\hat{\beta}T| \geq 1$. Then
$$
1+O(\lambda) \leq 
\left|\dis\frac{\ell_k(T^{-1})}{\ell_k(\hat{\beta})}\right|,
\left|\frac{\lambda_k(\hat{\beta})}{\lambda_k(T^{-1})}\right| 
\leq 1+O(\lambda)(1+\log|\hat{\beta}T|)
$$
\item[(7)]
$\left|\hat{\beta} \ \dis\frac{d \ln \ \ell(\hat{\beta})}{d\hat{\beta}}\right| \leq c_2|\lambda_{\hat{\beta}}|$.
\item[(8)]
$\left| \ln \ \dis\frac{\ell_k(\hat{\beta})}{\ell(\hat{\beta})}\right| \leq O(\lambda_k)(1+\log(1+ 
|\hat{\beta}_k|^{-1}))$.
\end{itemize}
\end{proposition}

This proposition plays a role in the proofs of Theorem \ref{theorem1.2} and of Theorem II.1.1. It will be proven along with Proposition \ref{proposition1.3} in Section 4. It turns out that we can use $\beta_{{\rm eff},\infty}$ in place of $\beta_{{\rm eff},N(x)}$ in Theorem II.1.1, as the following result shows (see Section 2 for a proof).

\begin{corollary}\label{corollary1.6}
Under the same assumptions as in Theorem II.1.1,
\begin{equation}\label{equation1.19}
|G_\lambda(\beta,x)-G_0(\beta_{{\rm eff},\infty},x)| \leq O(\lambda_{N(x)})|G_0(\beta_{{\rm eff},\infty},x)|.
\end{equation}
\end{corollary}

\subsection{Additional remarks}

In this paper we use a strategy of analyzing inverse Laplace transforms in order to obtain asymptotics as $T \rarrow \infty$. As a by-product we find it necessary to prove the needed Green's function estimates throughout a sector of the complex $\beta$-plane. for some models, it may be inconvenient to have complex coupling constants, so a natural question to ask is whether there are other ways of relating the asymptotics as $\beta$ tends to zero to the asymptotics as $T \rarrow \infty$. Tauberian theorems \cite{F} provide one answer, albeit a limited one. Working on the real axis, one can show that if $G(\beta)$ is the Laplace transform of a measure $\mu$, and $G$ varies regularly at 0, then $\mu\{[0,T]\}$ varies regularly at infinity and has an asymptotic behavior dual to the behavior of $G$ near zero. So, for example, if $G(\beta) \sim \beta^{-a}(\log \beta)^b$, then $\mu\{[0,T]\} \sim T^a(\log T)^b$.

The first problem we encounter is that in the hierarchical model, none of the quantities we work with behave regularly as $\beta \rarrow 0$ or as $T \rarrow \infty$. We need only look at Proposition \ref{proposition1.1} to see the type of behavior characteristic of a hierarchical model: asymptotically periodic in log $T$ or log $\beta$. One could perhaps get around this feature and prove a Tauberian theorem tailored to this situation, or work in a non-hierarchical model. However, there is still the problem of relating the asymptotics of $\mu$ to the asymptotics of the end-to-end distance. Tauberian theorems really only relate one type of average (the Laplace transform) to another $(\mu\{[0,T]\})$. To obtain results about the fixed $T$ ensemble of walks, one needs to learn about the density for $\mu$. In the situation at hand, $E^T_{0,\lambda}(|\omega(T)|^\alpha)$ is actually a ratio of two quantities, 
$\int dx \, P_\lambda(T,x)|x|^\alpha$ and $\int dx \, P_\lambda(T,x)$. These are inverse Laplace transforms of 
$$
\int dx \, G_\lambda(\beta,x)|x|^{\tilde{\alpha}} \sim (\beta \ell(\beta)^{-\frac{1}{4}})^{-1-\tilde{\alpha}/2} \mbox{ with } \tilde{\alpha} = \alpha \mbox{ or } 0, \mbox{ respectively.}
$$
Thus while the measures behave as $(T\ell(T^{-1})^{\frac{1}{4}})^{1+\tilde{\alpha}/2}$, we need to know that the densities behave as $1/T$ times this, or $T^{\tilde{\alpha}/2} \ell(T^{-1})^{(2+\tilde{\alpha})/8}$. Only with this information can we take the ratio and deduce that
$$
E^T_{0,\lambda}(|\omega(T)|^\alpha) \sim (T^{\frac{1}{2}} \ell(T^{-1})^{\frac{1}{8}})^\alpha,
$$
as described in Theorem \ref{theorem1.2}.
Without further assumptions, such as monotonicity, one cannot conclude much about the density knowing only the behavior of the measure. One can say that if the density has reasonable asymptotics as $T \rarrow \infty$, then they follow that of $\mu$. It should be clear, however, that working in the complex plane provides the most complete picture of the relation between the Green's function and the end-to-end distance. 

\bigskip

{\bf Related work.} Iagolnitzer and Magnen \cite{IM} have given detailed estimates on the decay of the critical Green's function for the Edwards model of weakly self-repelling polymers in four dimensions. Golowich \cite{G} extended their method into the region $\calD_\beta \setminus \calB(\varepsilon)$ with $\varepsilon > 0$.  Hara and Slade \cite{HS} have proved that the strictly self-avoiding walk on a simple cubic lattice $\mathbb{Z}^d$ for $d \geq 5$ has an end-to-end distance that is asymptotic to a constant times $\sqrt{T}$ and a scaling limit that is Brownian motion.  Golowich and Imbrie \cite{GI} obtained results on the critical behavior of the broken phase ($\beta < \beta^c(\lambda)$) of the hierarchical self-avoiding walk in four dimensions. Hattori and Tsuda \cite{HT} have detailed results on self-avoiding walks on the Sierpi\'{n}ski gasket.

\section{End-to-End Distance for the Non-interacting Walk}
\setcounter{equation}{0}

In this section we prove Proposition \ref{proposition1.4} (behavior of $G_0$) and then use the Laplace inversion formula to obtain the end-to-end distance and prove Proposition \ref{proposition1.1}. We also establish Corollary \ref{corollary1.6}.
\bigskip

{\em Proof of Proposition \ref{proposition1.4}}. From (2.15) of \cite{BEI} we have the following formula for $d=4$:
\begin{equation}\label{equation2.1}
G_0(\beta,x) = \sum_{k\geq 0} L^{-2k} \frac{1}{1+L^{2k}\beta}
\left(\boldone_{\{|x/L^k|=0\}}-L^{-4} \boldone_{\{|x/L^k|\leq L\}}\right).
\end{equation}
For $x \neq 0$ this can be written as
\begin{equation}\label{equation2.2}
G_0(\beta,x) = \sum_{k\geq N-1} L^{-2k} \frac{1-L^{-4}}{1+L^{2k}\beta}-
L^{-2(N-1)} \frac{1}{1+L^{2(N-1)}\beta},
\end{equation}
where $N = N(x) = \log|x|$. (Recall that $x \rarrow x/L$ means shifting the components of $x$ so that $x/L \equiv (\ldots,0,0,x_{N-1},x_{N-2},\ldots,x_1)$.)

We manipulate this expression in order to manifest cancellations between the two terms. Writing $\frac{1}{1+a} = \frac{1}{a} - \frac{1}{a(1+a)}$ with $a = L^{2k}\beta$ and using $\Sigma L^{-4k}(1-L^{-4}) = 1$ twice, we obtain
\begin{eqnarray*}
G_0(\beta,x)
& = & - \sum_{k \geq N-1} L^{-4k} \frac{(1-L^{-4})}{\beta(1+L^{2k}\beta)} + L^{-4(N-1)} \frac{1}{\beta(1+L^{2(N-1)}\beta)}  \\[4mm]
& = & \sum_{k \geq N-1} L^{-4k}(1-L^{-4}) \left(\frac{1}{\beta(1+L^{2(N-1)}\beta)} - \frac{1}{\beta(1+L^{2k}\beta)}\right).
\end{eqnarray*}
Clearing denominators and using $|x| = L^N$, $j = k-N$, we obtain
\begin{eqnarray}\label{equation2.3}
G_0(\beta,x)
& = &  \sum_{k \geq N-1} L^{-4k}(1-L^{-4}) 
\frac{(L^{2k}-L^{2(N-1)})}{(1+L^{2(N-1)}\beta)
(1+L^{2k}\beta)} \nonumber \\[4mm]
& = & \sum_{j \geq 0} 
L^{-2j}\frac{(1-L^{-4})(1-L^{-2-2j})}
{|x|^2(1+\beta|x|^2L^{-2})(1+\beta|x|^2L^{2j})}  \\[4mm]
& = & \frac{(1-L^{-4})(1-L^{-2})}{|x|^2(1+\beta|x|^2L^{-2})(1+\beta|x|^2)}
\left(
1+\sum_{j\geq 1} L^{-2j} \frac{1-L^{-2-2j}}{1-L^{-2}} \
\frac{1+\beta|x|^2}{1+\beta|x|^2L^{2j}}\right), \nonumber 
\end{eqnarray}
which leads to Proposition \ref{proposition1.4}(1).
For (2) we set $x=0$ in (\ref{equation2.1}):
\begin{eqnarray}\label{equation2.4}
G_0(\beta,0) 
& = &  \sum_{k \geq 0} L^{-2k} \ \frac{1-L^{-4}}{1+L^{2k}\beta} \nonumber \\[4mm]
& = & \frac{1-L^{-4}}{1+\beta} \left(1+ \sum_{k \geq 1} L^{-2k} \frac{1+\beta}{1+L^{2k}\beta}\right).
\end{eqnarray}

Proceeding to (3), we bound (\ref{equation2.3}) from above, noting that any $\chi \in \overline{\calD}_\beta$ has $|\mbox{arg } \chi| < b_\beta + \frac{1}{4} \, b_\lambda + \epsilon < \frac{3\pi}{4}$ and hence satisfies $|1+\chi| > 2^{-1/2}$. Thus, both $(1+\beta|x|^2)$ and $(1+\beta|x|^2L^{-2})$ are bounded below by $c^{-1}L^{-2}(1+|\beta| \, |x|^2)$, and in addition,
$$
\left|\frac{1+\beta|x|^2}{1+\beta|x|^2L^{2j}}\right| \leq c,
$$
uniformly in $\beta$, $|x|$, $j$, $L$. Hence the sum on $j$ converges, and the desired bound $c_2L^2|x|^{-2}(1+\beta|x|^2)^{-2}$ as in (3) follows. For the lower bound, we need only observe that for each $j$, $\mbox{arg}(1+\beta|x|^2L^{2j})$ lies between $\mbox{arg}(1+\beta|x|^2)$ and $\mbox{arg } \beta$ (and all three have the same sign). Hence each factor $(1+\beta|x|^2)/(1+\beta|x|^2L^{2j})$ is in $\overline{\calD}_\beta$ and on the same side of the real axis. So any positive linear combination of these factors is in $\overline{\calD}_\beta$. Using again the fact that $|1+\chi| \geq 2^{-1/2}$ for any 
$\chi \in \overline{\calD}_\beta$, we obtain a lower bound of the same form as the upper bound.  Similar arguments can be applied to the second line in (2.4), and the desired bounds on $G_0(\beta,0)$ follow. \hspace*{20pt} $\Box$

We need to control derivatives of $G_0(\beta,x)$ as well.

\begin{proposition}\label{proposition2.1}
If $\beta \in \overline{\calD}_\beta$, then for $x \neq 0$,
\begin{equation}\label{equation2.5}
\left|\beta \, \frac{d}{d\beta} \, G_0(\beta,x)\right| \leq \frac{cu(1+\log(1+u^{-1}))}{|x|^2(1+u)^3},
\end{equation}
where $u = |\beta| \, |x|^2$. For $x=0$, put $v = |\beta|$ and then
\begin{equation}\label{equation2.6}
\left|\beta \, \frac{d}{d\beta} \, G_0(\beta,0)\right| \leq \frac{cv(1+\log(1+v^{-1}))}{(1+v)^2}.
\end{equation}
\end{proposition}

Note that (\ref{equation2.5}) improves the naive bound $c|x|^{-2}(1+u)^{-2}$ that would follow from Proposition \ref{proposition1.4}(3). This is possible because the Green's function is relatively insensitive to changes in $\beta$ for smaller values of $|x|$.

{\em Proof.}  Consider what happens when $\beta \, \frac{d}{d\beta}$ is applied to the right-hand side of Proposition \ref{proposition1.4}(1). Wherever the derivative acts,
a new factor $\frac{uL^{2j}}{1+uL^{2j}}$ appears after taking absolute values. When $j=-1$, this is a constant times $\frac{u}{1+u}$ times our previous estimate, $c|x|^{-2}(1+u)^{-2}$. For $j \geq 0$, the $L^{-2j}$ which previously controlled the sum on $j$ is cancelled out, leaving a bound
$$
\sum_{j\geq 0} \frac{cu}{(1+u)(1+uL^{2j})^2}.
$$
If $u > 1$ this is still a geometric series, but for $u < 1$ there are $O(1+\log(1+u^{-1}))$ terms of approximately the same magnitude before convergence sets in, and this leads to the form of the bound (\ref{equation2.5}).

The same steps can be applied when estimating $\beta \, \frac{d}{d\beta} \, G_0(\beta,0)$. 
Differentiation of (\ref{equation2.4}) yields
$$
\left| \beta \, \frac{d}{d\beta} \, G_0(\beta,0)\right| \leq \sum_{j \geq 0} \frac{c}{(1+vL^{2j})^2},
$$
and proceeding as above we obtain (\ref{equation2.6}), and the proof is complete.
\hspace*{20pt} $\Box$
\bigskip

{\em Proof of Corollary \ref{corollary1.6}}. Use the bound from Proposition II.6.1,
$$
|G_\lambda(\beta,x)-G_0(\beta_{{\rm eff},N(x)},x)| \leq O(\lambda_{N(x)})|G_0(\beta_{{\rm eff},N(x)},x)|.
$$
Consider first $x \neq 0$ and let $N = N(x)$. We may apply Proposition \ref{proposition1.4} to the right-hand side. Proposition \ref{proposition1.5}(4) shows that $|\ell_k(\hat{\beta})|$ is essentially an increasing function of $k$.  Hence $|\ell_N(\hat{\beta})| \leq c|\ell(\hat{\beta})|$, so that $|\beta_{{\rm eff},N}| \geq c^{-1}|\beta_{{\rm eff},\infty}|$ and
\begin{equation}\label{equation2.7}
\frac{O(\lambda_N)}{|x|^2(1+|\beta_{{\rm eff},N}| \, |x|^2)^2} \leq
\frac{O(\lambda_N)}{|x|^2(1+|\beta_{{\rm eff},\infty}| \, |x|^2)^2} \leq O(\lambda_N)|G_0(\beta_{{\rm eff},\infty},x)|.
\end{equation}
We also need to estimate
\begin{eqnarray*}
|G_0(\beta_{{\rm eff},N},x)-
G_0(\beta_{{\rm eff},\infty},x)|
& = & \left|
\int^{\beta_{{\rm eff},\infty}}_{\beta_{{\rm eff},N}} \frac{d\tilde{\beta}}{\tilde{\beta}}
\, \tilde{\beta} \, \frac{d}{d\tilde{\beta}} \, G_0(\tilde{\beta},x)\right| \\[4mm]
& \leq &   
O(\lambda_N)(1+\log(1+|\hat{\beta}_N|^{-1}))
\sup_{\tilde{\beta}} \frac{u_{\tilde{\beta}}(1+\log(1+u_{\tilde{\beta}}^{-1}))}{|x|^2(1+u_{\tilde{\beta}})^3},
\end{eqnarray*}
where we have used Proposition \ref{proposition1.5}(8) and (\ref{equation2.5}) and put $u_{\tilde{\beta}}=|\tilde{\beta}||x|^2$. 
Let $u_N = |\hat{\beta}_N| = |\beta_{{\rm eff},N}||x|^2 \geq c^{-1}u_{\tilde{\beta}}$.  Assuming $u_N < 1$, we can use monotonicity to replace $u_{\tilde{\beta}}$ with $u_N$ in the sup.  The result is
$$
|G_0(\beta_{{\rm eff},N},x)-G_0(\beta_{{\rm eff},\infty},x)| \leq
\frac{O(\lambda_N)}{|x|^2(1+u_N)^2} \ \cdot \
\frac{u_N(1+\log(1+u_N^{-1}))^2}{1+u_N}.
$$
The second factor on the right-hand side is uniformly bounded, and the first factor is bounded by (\ref{equation2.7}). 
If $u_N \geq 1$, then $\log(1+|\hat{\beta}_N|^{-1}) \leq c$,
$u_{\tilde{\beta}}(1+u_{\tilde{\beta}})^{-1}(1+\log(1+u_{\tilde{\beta}}^{-1})) \leq c$, and $(1+u_{\tilde{\beta}})^{-2} \leq (1+c^{-1}|\beta_{{\rm eff},\infty}| |x|^2)^{-2}$, so we are still able to obtain the bound of (\ref{equation2.7}).
This establishes (\ref{equation1.19}) for $x \neq 0$.

The case $x=0$ can be handled similarly. When (\ref{equation2.6}) is combined with 
$$
\left|\ln{\frac {\beta}{\beta_{{\rm eff},\infty}}}\right| \leq O(\lambda)(1+\log(1+v^{-1}))
$$
as above ({\em c.f.} Proposition \ref{proposition1.5}(8) with $k=0$), we obtain (\ref{equation1.19}). This completes the proof.
\hspace*{20pt} $\Box$
\bigskip

{\em Proof of Proposition \ref{proposition1.1}.} Let $P_0(T,x)$ be the transition probability for the L\'evy process. From the definition of $G_0(\beta,x)$ and the Laplace transform inversion formula, we have
\begin{equation}\label{equation2.8}
P(T,x) = \int \frac{d\beta}{2\pi i} \ e^{\beta T} G_0(\beta,x),
\end{equation}
where the contour is $\{\beta: \ \beta = a+i\alpha, \ \alpha \in \mathbb{R}, \ a > 0\}$. We can move the contour to the left and close it so that it encircles the poles in $[-1,0)$, {\em c.f.}~(\ref{equation2.2}). By interchanging the integral over $\beta$ with the sum in (\ref{equation2.2}) and applying the residue formula, we obtain for $x = L^N$, $N \geq 1$,
$$
P_0(T,x) = \sum_{k \geq N-1} L^{-4k}(1-L^{-4})e^{-L^{-2j}T}-L^{-4(N-1)}e^{-2(N-1)T}.
$$
Using $j = k-N$ and $\sum L^{-4j}(1-L^{-4}) =1$, this becomes
\begin{eqnarray}\label{equation2.9}
P_0(T,x) & = & L^{-4N} \sum_{j \geq 0} L^{-4j}(1-L^{-4})\left(e^{-L^{-2j}L^{-2N}T}-e^{-L^{-2(N-1)}T}\right)
\nonumber \\[4mm]
& = & |x|^{-4}f(t),
\end{eqnarray}
where $t = T/|x|^2$ and
\begin{equation}\label{equation2.10}
f(t) = \sum_{j \geq 0} L^{-4j}(1-L^{-4})\left(e^{-L^{-2j}t}-e^{-L^{2}t}\right).
\end{equation}

The following proposition gives an accurate picture of the shape of $P_0(T,x)$.

\begin{proposition}\label{proposition2.2}
Let $x \neq 0$. Then there are constants $c_1, c_2$ such that
\begin{equation}\label{equation2.11}
\frac{c_1}{T^2\left(1+ \frac{|x|^2}{T}\right)^3} \leq P_0(T,x) \leq \frac{c_2}{T^2\left(1+ \frac{|x|^2}{T}\right)^3}.
\end{equation}
This estimate holds also for $x=0$, provided $T \geq 1$. For small $T$, $P_0(T,0) \sim 1-O(T)$.
\end{proposition}

{\em Proof.}
Note that for $t < 1$, $f(t) \sim t$. For $t > 1$, the sum is (\ref{equation2.10}) is dominated by the term with $L^{-2j}t \approx 1$, and so $f(t) \sim t^{-2}$. Overall, $f(t)$ is bounded above and below by positive multiples of $t^{-2}(1+t^{-1})^{-3}$, which implies (\ref{equation2.11}).
To handle the case $x=0$, we use Proposition \ref{proposition1.4}(2) and (\ref{equation2.8}) to obtain
$$
P_0(T,0) = \sum^\infty_{k=0} L^{-4k}(1-L^{-4}) e^{-L^{-2k}T},
$$
which behaves as $T^{-2}$ for $T \geq 1$ and $1-O(T)$ for $T < 1$. Thus (\ref{equation2.11}) holds for $x=0$, provided $T \geq 1$. \hspace*{20pt} $\Box$
\bigskip

Continuing with the proof of Proposition \ref{proposition1.1}, note that from
(\ref{equation2.9}), for $0 < \alpha <2$, we have
\begin{eqnarray}\label{equation2.12}
E_0\left(\frac{|\omega(T)|^\alpha}{T^{\alpha /2}}\right)
& \equiv & \int dx \, P_0(T,x) \, \frac{|x|^\alpha}{T^{\alpha/2}} \nonumber \\[4mm]
& = & \sum_{N\geq 1} L^{4N}(1-L^{-4})P_0(T,x)|_{|x| = L^N}
\frac{L^{\alpha N}}{T^{\alpha/2}} \nonumber \\[4mm]
& = & \sum_{N \geq 1} f_\alpha\left(\frac{T}{L^{2N}}\right),
\end{eqnarray}
where $f_\alpha(t) = t^{-\alpha/2}(1-L^{-4})f(t)$. Now we replace $T$ by $L^{2m}T$ in (\ref{equation2.12}) and find that as $m \rarrow \infty$,
$$
E_0 \left(\frac{|x|^\alpha}{(L^{2m}T)^{\alpha/2}}\right) = \sum_{j \geq 1-m} f_\alpha(T/L^{2j}) \rarrow \sum^\infty_{j=-\infty} f_\alpha(T/L^{2j}).
$$
Since $f_\alpha(t)$ goes to zero at 0 and $\infty$ as a power of $t$, the sum on $j$ converges at both ends and defines a function with the properties claimed in Proposition \ref{proposition1.1}. \hspace*{20pt} $\Box$

\section{End-to-End Distance for the Self-Avoiding Walk}
\setcounter{equation}{0}

We begin with a detailed statement of the behavior of the (unnormalized) transition probability function for the interacting model. Let
\begin{equation}\label{equation3.1}
P_\lambda(T,x) \equiv E_0 \left(e^{-\lambda \tau^{2}(\calG)-\beta^{c}(\lambda)T}\boldone_{\{ \omega(T)=x\}}\right).
\end{equation}
Then $G_\lambda(\beta,x)$ is the Laplace transform of $P_\lambda(T,x)$, so as in (\ref{equation2.8}) we have
\begin{eqnarray}\label{equation3.2}
P_\lambda (T,x) & = & \int \frac{d\beta}{2\pi i} \ e^{(\beta-\beta^{c}(\lambda))T}G_\lambda(\beta,x) \nonumber \\[4mm]
& = & \int \frac{d\hat{\beta}}{2\pi i} \ e^{\hat{\beta}T}G_\lambda(\beta,x),
\end{eqnarray}
where $\hat{\beta} = \beta -\beta^c(\lambda)$. In this equation we
may, by Theorem II.1.1 { and Proposition~\ref{proposition1.4}, } choose the contour to be $T^{-1}\Gamma$, where $\Gamma$ consists of the two rays $\{z: |z| \geq 1 \mbox{ and arg }z = \pm b_\beta\}$ joined by an arc of the unit circle which passes across the positive real axis. Recall that $\frac{\pi}{2} < b_\beta < \frac{3\pi}{4}$ and that $b_\lambda < \pi-\frac{4}{3}b_\beta < \frac{\pi}{3}$.

\begin{proposition}\label{proposition3.1}
Let $k = \max\{0,\log|x|\}$ and put $\hat{\beta} = T^{-1}$ in $\lambda_k = \lambda_k(T^{-1})$. Likewise, define $\ell = \ell(T^{-1})$, where $\ell(\hat{\beta}) = (\beta_{{\rm eff},\infty}/\beta)^{-4}$ as per Proposition \ref{proposition1.5}. Then with $t \equiv T\ell^{\frac{1}{4}} > 1$, the following estimate holds uniformly in $x, T$ and $\lambda \in \calD_\lambda$:
\begin{eqnarray}\label{equation3.3}
P_\lambda(T,x) & = & \ell^{\frac{1}{4}} \left(P_0(t,x) + \frac{O(\lambda_k)}{t(1+|x|^2)\left(1+ |x|^2/t\right)^2}\right) \nonumber \\[4mm]
& = & \ell^{\frac{1}{4}} P_0(t,x) \left(1+O(\lambda_k)\left(\frac{t+|x|^2}{1+|x|^2}\right)\right).
\end{eqnarray}
\end{proposition}

{\em Proof.} Corollary \ref{corollary1.6} estimates $G_\lambda$ in terms of $G_0$:
$$
G_\lambda(\beta,x) = G_0(\beta_{{\rm eff},\infty}(\hat{\beta}),x)(1+O(\lambda_k(\hat{\beta}))).
$$
We need to replace $\hat{\beta}$ with $T^{-1}$ in part of this expression.  To simplify formulas, let us put
$$
G_0(\zeta) = G_0(\hat{\beta} \ell(\zeta)^{-\frac{1}{4}},x).
$$
so that $G_0(\beta_{{\rm eff},\infty}(\hat{\beta}),x) = G_0(\hat{\beta})$.  Then we have
\begin{eqnarray*}
P_\lambda(T,x)
& = & \int \frac{d\hat{\beta}}{2\pi i} \, e^{\hat{\beta}T} \left[G_0(T^{-1})+O(\lambda_k(\hat{\beta}))G_0(\hat{\beta})+(G_0(\hat{\beta})-G_0(T^{-1}))\right]
\\[4mm]
& = & \int \frac{d\hat{\beta}}{2\pi i} \, e^{\hat{\beta}T} \left[G_0(
\hat{\beta} \ell^{-\frac{1}{4}},x)+\hat{e}_1(x,\hat{\beta}) + \hat{e}_2(T,x,\hat{\beta})\right]
\\[4mm]
& = & \ell^{\frac{1}{4}} \int \, \frac{d\beta'}{2\pi i} e^{\beta't}G_0(\beta',x)+e(T,x) 
\\[4mm]
& = & \ell^{\frac{1}{4}} P_0(t,x) +e(T,x),
\end{eqnarray*}
where $\hat{e}_1 = O(\lambda_k(\hat{\beta}))G_0(\hat{\beta})$, $\hat{e}_2 = G_0(\hat{\beta})-G_0(T^{-1})$, and $e(T,x)$ is the inverse Laplace transform of their sum. We have
\begin{eqnarray}\label{equation3.4}
|\hat{e}_2(T,x,\hat{\beta})|
& = & \left|\int^{\hat{\beta}}_{T^{-1}} d \tilde{\beta} \frac{d}{d\tilde{\beta}} \, G_0(\tilde{\beta})\right| \nonumber \\[4mm]
& = & \left|\int^{\hat{\beta}}_{T^{-1}} \frac{d\tilde{\beta}}{\tilde{\beta}}
\left(- \, \frac{1}{4}\right) \left(\tilde{\beta} \, \frac{d}{d\tilde{\beta}} \, \ln \ell(\tilde{\beta})\right) w \, \frac{\partial}{\partial w} \, G_0(w,x)\right|,
\end{eqnarray}
where $w = \hat{\beta} \ell(\tilde{\beta})^{-\frac{1}{4}}$. Put $u = |w| \, |x|^2$.
Then if $x \neq 0$, Proposition \ref{proposition2.1} implies that
\begin{eqnarray*}
\left|w \, \frac{\partial}{\partial w} \, G_0(w,x)\right|
& \leq & \frac{cu(1+\log(1+u^{-1}))}{|x|^2(1+u)^3} \leq \frac{c}{|x|^2(1+u)^2} \\[4mm]
& = & \frac{c}{|x|^2(1+|\hat{\beta} \ell(\tilde{\beta})^{-\frac{1}{4}}| \, |x|^2)^2} \leq \frac{c'}{|x|^2(1+|\hat{\beta} \ell^{-\frac{1}{4}}| \, |x|^2)^2}
\end{eqnarray*}
where in the last step we have used Proposition \ref{proposition1.5}(6).
For $x=0$, this bound has to be replaced with $c'(1+|\hat{\beta} \ell^{-\frac{1}{4}}|)^{-1}$. 

Continuing under the assumption that $x \neq 0$, we use this bound and Proposition \ref{proposition1.5}(7) to estimate (\ref{equation3.4}) by
$$
|\hat{e}_2(T,x,\hat{\beta})|
\leq \frac{\left(\ln|\hat{\beta}T|+ \frac{3\pi}{4}\right)\sup_{\tilde{\beta}}O(\lambda_{\tilde{\beta}})}{|x|^2(1+|\hat{\beta} \ell^{-\frac{1}{4}}| \, |x|^2)^2}
\leq \frac{\left(\ln|\hat{\beta}T|+ \frac{3\pi}{4}\right)\sup_{\tilde{\beta}}O(\lambda_k(\tilde{\beta}))}{|x|^2(1+|t|^{-1} \, |x|^2)^2}.
$$
In the second inequality, we have used $|\tilde{\beta}| \geq T^{-1}$, $t=T\ell^{\frac{1}{4}}$, and the fact that $|\lambda_k(\tilde{\beta})|$ is essentially a decreasing function of $k$ ({\em c.f.}~Proposition \ref{proposition1.5}(3)).  Note that if we use Proposition \ref{proposition1.4} to estimate $G_0(\hat{\beta})$, we find that $\hat{e}_1(x,\hat{\beta})$ is bounded by this same expression, only with $O(\lambda_k(\tilde{\beta}))$ replaced by $O(\lambda_k(\hat{\beta}))$. Hence we combine the two error terms and estimate $|\lambda_k(\tilde{\beta})| \leq \lambda_k(1+O(\lambda)(1+\log|\tilde{\beta}T|)) \leq \lambda_k(1+\log|\hat{\beta}T|)$ ({\em c.f.}~Proposition \ref{proposition1.5}(6)) to obtain
$$
|e(T,x)| \leq \frac{O(\lambda_k)}{T|x|^2(1+|x|^2/t)^2} \int_\Gamma \left|d (\hat{\beta}T) e^{\hat{\beta}T}\right| \,
(1+\ln|\hat{\beta}T|)^2 
.
$$
As $e^{\hat{\beta}T}$ decays exponentially on the rays $|\mbox{arg }\beta| = b_\beta$, the integral is $O(1)$ and so
\begin{eqnarray}\label{equation3.5}
|e(T,x)|
& \leq & \frac{O(\lambda_k)}{T|x|^2\left(1+|x|^2/t\right)^2}
= \frac{O(\lambda_k)\ell^{\frac{1}{4}}}{t(1+|x|^2)\left(1+ |x|^2/t\right)^2},
\end{eqnarray}
which implies (\ref{equation3.3}).  The second statement in (\ref{equation3.3}) follows from this by using (\ref{equation2.11}), with $T$ replaced by $t=T\ell^{\frac{1}{4}}$.  We note that 
$$
|\mbox{arg }t|=|\mbox{arg }\ell^{\frac{1}{4}}|=\frac{1}{4}|\mbox{arg}(1+8B\lambda k_{1/T})|+O(\lambda)<\frac{1}{4}|\mbox{arg }\lambda|+O(\lambda)<\frac{\pi}{12}+O(\lambda)
$$
({\em c.f.} Proposition \ref{proposition1.5}(5)), and that the proof of Proposition \ref{proposition2.2} extends to the continuation of $P_0(T,x)$ into this sector.

The case $x=0$ is handled similarly, only $|x|$ has to be replaced with 1 and the power of $(1+|x|^2/t)$ is reduced from 2 to 1. The final bound in (\ref{equation3.5}) remains valid, however. \hspace*{20pt} $\Box$
\bigskip

{\bf Remark.} The error term in (\ref{equation3.3}) behaves as $t^{-1}(1+|x|^2)^{-1}$ for $|x|^2 < t$, which is not the behavior one would expect (namely $t^{-2}$, the small-$x$ behavior of $P_0(t,x)$). This is an artifact of the proof, which takes an absolute value of $G_0$ on the contour, thereby spoiling the cancellations needed to get a bound proportional to $t^{-2}$, and leading to ``Green's function-like'' rather than ``transition probability-like'' behavior. While (\ref{equation3.3}) is adequate for obtaining our main theorem on the end-to-end distance, it may be of some interest to indicate how a better bound might be proven. Let $\ell_k(\hat{\beta})$ be as in Proposition \ref{proposition1.5} and put
$\ell_k = \ell_k(T^{-1})$ and $t_k = T \ell_k^{\frac{1}{4}}$, with $k = \max\{0,\log|x|\}$ as in Proposition \ref{proposition3.1}. Then we conjecture
\begin{equation}\label{equation3.6}
P_\lambda(T,x) = \ell_k^{\frac{1}{4}} P_0(t_k,x)(1+O(\lambda_k)).
\end{equation}
To get this, one need only consider $|x|^2 < t_k$ as the arguments above give it for $|x|^2 \geq t_k$. Write
$$
P_\lambda(T,x) = \frac{1}{T^2} \int \frac{d \hat{\beta}}{2\pi i} \, e^{\hat{\beta}T} \, G''_\lambda(\beta,x),
$$
where primes denote $\hat{\beta}$-derivatives. One should be able to replace $G''_\lambda(\beta,x)$  with $G''_0(w,x)$ plus error terms of order $\lambda_k(\hat{\beta})G''_0(w,x)$, $\lambda'_k(\hat{\beta})G'_0(w,x)$, and $\lambda''_k(\hat{\beta})G_0(w,x)$, where $w=\beta\ell_k(\beta)^{-\frac{1}{4}}$. 
Each $\beta$-derivative of $G_0(w,x)$ is actually $\ell_k(\hat{\beta})^{-\frac{1}{4}}(1+O(\lambda_k(\hat{\beta})))$ times the corresponding $w$-derivative, 
the correction term being $\hat{\beta} \, \frac{d}{d\hat{\beta}} \, \ln \ell_k(\hat{\beta})$, which as in Proposition \ref{proposition1.5}(7), is $O(\lambda_k(\hat{\beta}))$.
Extending the proof of Proposition \ref{proposition2.1} to higher derivatives, we have
$$
\begin{array}{rclcrcl}
|G_0(w,x)| & \leq & \frac{c}{|x|^2(1+u)^2} = \frac{c|w|}{u(1+u)^2},
&
|\frac{\partial}{\partial w}G_0(w,x)| & \leq & \frac{c\log(1+u^{-1})}{(1+u)^3}, \\[4mm]
|\frac{\partial^2}{\partial w^2}G_0(w,x)| & \leq & \frac{c}{|w|(1+u)^3},
&
|\frac{\partial^3}{\partial w^3}G_0(w,x)| & \leq & \frac{c}{|w|^2(1+u)^3}.
\end{array}
$$
Extending the arguments of Lemma \ref{lemma4.2} to second derivatives, we expect
\begin{eqnarray*}
|\lambda'_k(\hat{\beta})| & \leq & O(\lambda_k(\hat{\beta})^2) \beta'_k = O(\lambda_k(\hat{\beta})^2) \hat{\beta}_k/\hat{\beta}, \\[4mm]
|\lambda''_k(\hat{\beta})| & \leq & O(\lambda_k(\hat{\beta})^3) {\beta'_k}^2 = O(\lambda_k(\hat{\beta})^3) \hat{\beta}^2_k/\hat{\beta}^2.
\end{eqnarray*}
We shall see that the factors of 
$|\beta_k|=|\hat{\beta}L^{2k}\ell_k(\hat{\beta})^{-\frac{1}{4}}|=|w| \, |x|^2= u$ 
in $\lambda'_k$, $\lambda''_k$ control the dangerous $u^{-1}$ and log$(1+u^{-1})$ factors in $G_0$, $G'_0$ respectively. Noting that $\hat{\beta}/w=\ell_k(\hat{\beta})^{\frac{1}{4}}$, we find that $\hat{e}_1(x,\hat{\beta})=G''_\lambda -G_0''$ is bounded by
\begin{eqnarray*}
&& O(\lambda_k(\hat{\beta}))\frac{\ell_k(\hat{\beta})^{-\frac{1}{2}}}{w(1+u)^3}
+O(\lambda_k(\hat{\beta})^2)\frac{u}{\hat{\beta}}\frac{\ell_k(\hat{\beta})^{-\frac{1}{4}}\log(1+u^{-1})}{(1+u)^3}+O(\lambda_k(\hat{\beta})^3)\frac{u^2}{\hat{\beta}^2}\frac{w}{u(1+u)^2}
\\[4mm]
& \leq & O(\lambda_k(\hat{\beta}))\hat{\beta}^{-1} \left(|\ell_k(\hat{\beta})|^{-\frac{1}{2}}\left|\frac{\hat{\beta}}{w}\right| + |\lambda_k(\hat{\beta})|\,|\ell_k(\hat{\beta})|^{-\frac{1}{4}} + |\lambda_k(\hat{\beta})|^2\left|\frac{w}{\hat{\beta}}\right|\right) \\[4mm]
& \leq &
O(\lambda_k(\hat{\beta}))\hat{\beta}^{-1}\ell_k(\hat{\beta})^{-\frac{1}{4}}  
\leq 
O(\lambda_k)\hat{\beta}^{-1}\ell_k^{-\frac{1}{4}} (1+\log|\hat{\beta}T|)^{\frac{5}{4}}.
\end{eqnarray*}
Furthermore, $e_2(T,x,\hat{\beta})$ satisfies the same bound because $\left|\tilde{\beta} \, \frac{d}{d\tilde{\beta}} \, \ln \ell_k(\tilde{\beta})\right|
\leq O(\lambda_k(\tilde{\beta}))$ and because the bound on $wG'''_0$ is the same as the one on $G''_0$.
One can perform the inverse Laplace transform on this and estimate it as in the proof of Proposition \ref{proposition3.1}. The result is $|e(T,x)| \leq O(\lambda_k)\ell_k^{-\frac{1}{4}}$, which when multiplied by $T^{-2} =(t_k\ell_k^{-\frac{1}{4}})^{-2}\approx \ell_k^{\frac{1}{2}}P_0(t_k,x)$ ({\em c.f.} Proposition \ref{proposition2.2}), leads to (\ref{equation3.6}).
\bigskip

{\em Proof of Theorem \ref{theorem1.2}.} By (\ref{equation3.3}), we have
\begin{eqnarray*}
\int dx \, P_\lambda(T,x)|x|^\alpha
& = & \ell^{\frac{1}{4}}
\left( \int dx \, P_0(t,x)x^\alpha + \sum_k
\ \frac{O(\lambda_k)(L^{4k}-1)L^{\alpha k}}{t(1+L^{2k})(1+L^{2k}/t)^2}\right)
\\[4mm]
& = & \ell^{\frac{1}{4}}\left(E_0(|\omega(t)|^\alpha)+O(\lambda_{k_{1/T}})t^{\alpha/2}\right)
\\[4mm]
& = & \ell^{\frac{1}{4}} E_0(|\omega(t)|^\alpha)\left(1+O(\lambda_{k_{1/T}})\right).
\end{eqnarray*}
Since $\lambda_k$ varies slowly with $k$ and $0 \leq \alpha < 2$, the sum on $k$ first increases geometrically, then decreases geometrically, so that the sum on $k$ is estimated by the largest term $k = \bar{k}$, for which $L^{2\bar{k}} \approx t$. We have replaced $\bar{k}$ with $k_{1/T}$, which is allowable because 
at $\hat{\beta} = T^{-1}$,
$$
\hat{\beta}_{\bar{k}} = \hat{\beta} L^{2\bar{k}} \ell(\hat{\beta})^{-\frac{1}{4}}|_{\hat{\beta}=T^{-1}} \approx T^{-1} t \ell^{-\frac{1}{4}} = 1,
$$
so that $\bar{k} \approx k_{1/T}$. Note that Proposition \ref{proposition1.5}(2) relates $k_{1/T}$ to $T$:
\begin{equation}\label{equation3.7}
k_{1/T} = O(1) + \frac{1}{2} \log(1+T) + \frac{1}{8} \log |1+4B\lambda \log(1+T)|.
\end{equation}
In fact, we can use Proposition \ref{proposition1.5}(3, 4, 8) to write
$$
\lambda_{k_{1/T}} \approx \frac{\lambda}{1+8B \lambda k_{1/T}} \approx \lambda \ell^{-1}_{k_{1/T}} \approx \lambda \ell^{-1}
$$
(equality to within a factor $e^{O(\lambda)})$. Hence
\begin{eqnarray}\label{equation3.8}
\int dx \, P_\lambda (T,x)|x|^\alpha 
& = & \ell^{\frac{1}{4}}(E_0(|\omega(t)|^\alpha)+O(\lambda \ell^{-1})t^{\alpha/2})
\nonumber \\[4mm]
& = & \ell^{\frac{1}{4}}E_0(|\omega(t)|^\alpha)(1+O(\lambda \ell^{-1})),
\end{eqnarray}
where we have used Proposition \ref{proposition1.1}. Using (\ref{equation3.8})
for numerator and denominator, we obtain
$$
E^T_{0,\lambda}(|\omega(T)|^\alpha) = E_0(|\omega(t)|^\alpha)(1+O(\lambda \ell^{-1})),
$$
which leads immediately to (\ref{equation1.7}).

We have $\ell = \ell(T^{-1}) = (1+8B \lambda k_{1/T}) e^{O(\lambda)}$, by Proposition \ref{proposition1.5}(5), and if we insert (\ref{equation3.7}) into this, we obtain (\ref{equation1.8}). \hspace*{20pt} $\Box$

\section{The Coupling Constant Recursion and its Fixed Point}
\setcounter{equation}{0}

This section begins with an inverse function theorem construction of the fixed point $\beta^c(\lambda)$, as specified in Proposition \ref{proposition1.3}. Then the shifted recursion for $\hat{\beta} = \beta - \beta^c(\lambda)$ is controlled in some detail, and Proposition \ref{proposition1.3} can be established.  Finally, these results are used to prove Proposition \ref{proposition1.5}.

As we shall see, one can prove accurate estimates on $\lambda_k$, $\beta_k$ by working inductively on domains which extend slightly into the ``dangerous'' region left of $\beta^c(\lambda)$.  Precise control of $\beta_k'$ is needed in order to obtain the right domain of analyticity for Cauchy estimates.  As $k \rightarrow \infty$, the domain shrinks back to $\calD_{\beta}+\beta^c(\lambda)$ as the singularity at $\beta^c(\lambda)$ asserts itself.  Proposition II.6.1 provides the necessary input.

We wish to construct $\beta^c(\lambda)$ as the limit of the decreasing sequence of open sets $\beta^{-1}_k(\calB(\frac{1}{2}))$. But we must show that the map $\beta_k(\beta)$ and its inverse are defined in appropriate domains. We establish the following lemma inductively (keep in mind that $\lambda$ is fixed in $\calD_\lambda$; $\lambda_k$ and $\beta_k$ are regarded as functions of $\beta$, with primes denoting $\beta$-derivatives).

We use the notation
\begin{equation}\label{equation4.1}
l_k = \left|\exp\left(\sum^{k-1}_{j=0} \, \frac{8B}{\lambda^{-1}+8 Bj}\right)\right|,
\end{equation}
and note that this is a function of $\lambda, k$ only. By integral approximation, it can easily be shown that $l_k = |1+8B\lambda k|e^{O(\lambda)}$.

\begin{lemma}\label{lemma4.1}Let $k\ge1$.

\begin{itemize}
\item[(1)]
$\beta_k$ and $\lambda_k$ are defined on $\beta_{k-1}^{-1}(\calB(\frac{1}{2}))$, and $(\beta_j,\lambda_j) \in \calB(\frac{1}{2}) \times \overline{\calD}_\lambda \mbox{ for } 0 \le j < k$.
\item[(2)]
$\lambda_k \in \overline{\calD}_\lambda$. Furthermore,
$$
\left|\lambda_k - \frac{1}{\lambda^{-1}+8Bk}\right| \leq c_1\left|\frac{1}{\lambda^{-1}+8Bk}\right|^2(1+\ln(1+|\lambda| k)),
$$
with $c_1$ a constant independent of $k$ and $\beta \in \beta_{k-1}^{-1}(\calB(\frac{1}{2}))$.
\item[(3)]
For $\beta \in \beta_{k-1}^{-1}(\calB(\frac{1}{3})), \ |\lambda_k'| \leq c_2|\lambda^2_k \beta'_k|$ and $|\beta'_k| = L^{2k} l_k^{-\frac{1}{4}}e^{O(\lambda)}$.  Here $O(\lambda)$ denotes a quantity bounded by $c_3|\lambda|$, and $c_2$, $c_3$ are independent of $k$ and $\beta$.
\item[(4)]
$\beta_k^{-1}$ is well-defined on $\calB(\frac{1}{2})$ and $\beta_k^{-1}(\calB(\frac{1}{2})) \subset \beta_{k-1}^{-1}(\calB(\frac{1}{3})) \subset \calB(\frac{1}{2})$.
\end{itemize}
\end{lemma}

{\em Proof.} Assume (1)--(4) up through $k$ and prove them for $k+1$. For the first case ($k+1=1$) we shall need only the bound of (3) for $k=0$, and this follows from the fact that $\lambda_0'=0$, $\beta_0'=1$.
In order to prove (1), work on $\beta_k^{-1}(\calB(\frac{1}{2}))$, defined by virtue of (4).  As (4) also implies
$\beta_k^{-1}(\calB(\frac{1}{2})) \subset \beta_{k-1}^{-1}(\calB(\frac{1}{2}))$, we may use (1) to put $(\beta_j,\lambda_j)$ in $\calB(\frac{1}{2}) \times \overline{\calD}_\lambda$ for $j<k$.  As (2) places $\lambda_k$ in
$\overline{\calD}_\lambda$, and as $\beta_k(\beta_k^{-1}(\calB(\frac{1}{2}))) = \calB(\frac{1}{2})$, we have the needed statement for 
$j=k$ as well.
Hence $(\beta_{k+1},\lambda_{k+1})$ is defined, by Proposition II.6.1, and it satisfies (\ref{equation1.16}), (\ref{equation1.17}).

We may rewrite the $\lambda$ recursion as
$$
\lambda^{-1}_{j+1} = \lambda_j^{-1} + \frac{8B}{(1+\beta_j)^2} + O(\lambda_j),
$$
where we have used the fact that $\beta_j \in \calB(\frac{1}{2})$ for all $0 \leq j < k$ to avoid writing some $(1+\beta_j)^{-1}$ factors. This implies that
\begin{equation}\label{equation4.2}
\lambda_{k+1}^{-1} = \lambda^{-1} + \sum^k_{j=0} \left[\frac{8B}{(1+\beta_j)^2} + O(\lambda_j)\right] = \lambda^{-1}+8B(k+1) +O(1)(1+\ln(1+|\lambda|(k+1))),
\end{equation}
where we have used
$$
\sum^k_{j=0} \left|\frac{1}{(1+\beta_j)^2} -1\right| = \sum^k_{j=0} O(\beta_j) \leq O(\beta_k) + \sum^k_{j=0} O(\lambda_j),
$$
$$
\sum^k_{j=0} O(\lambda_j) \leq O(1) \sum^k_{j=0} \left|\frac{1}{\lambda^{-1}+8Bj}\right| \leq O(1) \ln(1+|\lambda|(k+1)).
$$
The first of these bounds follows by bounding separately the set of $j$'s such that $|\beta_j| > |\lambda_j|$. Once this inequality holds, it holds for all larger $j$'s (with geometric growth of $\beta_j$) as is clear from (\ref{equation1.16}). The second bound follows from (2), keeping in mind that $\calD_\lambda$ is contained in a sector which does not include the negative reals, so $\lambda^{-1}$ and $8Bk$ never come close to canceling. Using the identity $\lambda - \tilde{\lambda} = \lambda \tilde{\lambda}(\tilde{\lambda}^{-1}-\lambda^{-1})$ we have
\begin{equation}\label{equation4.3}
\left|\lambda_{k+1} - \frac{1}{\lambda^{-1} +8B(k+1)}\right| = \lambda_k\left|\frac{1}{\lambda^{-1}+8B(k+1)}\right| O(1)(1+\ln(1+|\lambda|(k+1))),
\end{equation}
and the bound in (2) follows for $\lambda_{k+1}$. 

We now prove that $\lambda_k \in \overline{\calD}_\lambda$.  Note that if $\delta$ (which defines the maximum $|\lambda|$ in $\calD_\lambda$) is chosen small enough, then $|\lambda_{k+1}| \leq \overline{\delta}$. The sequence $\tilde{\lambda}_j = (\lambda^{-1}+8Bj)^{-1}$ follows a circle tangent to the real axis at 0, so that $|\mbox{arg } \tilde{\lambda}_j|$ is decreasing in $j$. Furthermore, the bound in Lemma \ref{lemma4.1}(2) shows that any increase in $|\mbox{arg }\lambda_j|$ in the exact recursion is at most $O(\lambda)$.
 Thus, while $\lambda_k$ may leave $\calD_\lambda$, it remains in $\overline{\calD}_\lambda$. We have now established (1) and (2).

To check (3), differentiate (\ref{equation1.16}):
\begin{eqnarray}
\lambda'_{k+1} & = & \lambda'_k - \, \frac{16B(\lambda_k\lambda'_k-\lambda^2_k(1+\beta_k)^{-1} \beta'_k)}{(1+\beta_k)^2} + \epsilon'_{\lambda,k}, \label{equation4.4} \\[4mm]
\beta'_{k+1} & = & L^2\left[\beta'_k + \, \frac{2B(\lambda'_k-\lambda_k (1+\beta_k)^{-1} \beta'_k)}{1+\beta_k}\right]+ \epsilon'_{\beta,k}. \label{equation4.5}
\end{eqnarray}
By the $\beta'_k$ bound in (3), the domain $\beta_k^{-1}(\calB(\frac{1}{2}))$ includes balls of size $\frac{1}{6} L^{-2k} l_k^{\frac{1}{4}}$. Hence, (\ref{equation1.17}), Cauchy's bound, and (3) imply
\begin{eqnarray}
|\epsilon'_{\lambda,k}|
& \leq & c|\lambda_k|^3|1+\beta_k|^{-1} L^{2k} l_k^{-\frac{1}{4}} \leq c|\lambda_k|^3|1+\beta_k|^{-1}|\beta'_k|,  \label{equation4.6} \\[4mm]
|\epsilon'_{\beta,k}|
& \leq & c|\lambda_k|^2|1+\beta_k|^{-2} L^{2k} l_k^{-\frac{1}{4}} \leq c|\lambda_k|^2|1+\beta_k|^{-2}|\beta'_k|, \label{equation4.7}
\end{eqnarray}
for $\beta \in \beta_k^{-1}(\calB(\frac{1}{3}))$. Inserting the bound (4.7) into (4.5) and using (3), we obtain
\begin{equation}\label{equation4.8}
\beta'_{k+1} = L^2 \beta'_k[1-2B \lambda_k+O(\beta_k \lambda_k)+O(\lambda_k^2)],
\end{equation}
which can be written in exponential form:
\begin{equation}\label{equation4.9}
\beta'_{k+1} = L^{2(k+1)} \exp\left[\sum^k_{j=0}(-2B \lambda_j+O(\beta_j \lambda_j)+O(\lambda^2_j))\right].
\end{equation}
Replacing $\lambda_j$ with $\lambda^{-1} +8Bj$ as per (2), we pick up an error $\sim \lambda^2_j(1+\ln(1+|\lambda|j))$, which, however, is summable in $j$. The other terms in (\ref{equation4.9}) also sum to $O(\lambda)$, so the $\beta'_{k+1}$ bound in (3) follows.

Moving on to the $\lambda'_{k+1}$ bound, we insert (3) into (\ref{equation4.4}):
\begin{equation}\label{equation4.10}
|\lambda'_{k+1}| \leq |\lambda^2_k \beta'_k|(c_2+O(\lambda_k)+O(1)),
\end{equation}
where $\epsilon'_{\lambda,k}$ has been bounded using (\ref{equation4.6}). Now, provided $c_2$ is chosen large enough, so that $L^{-2} e^{O(\lambda_k)}(c_2+O(1)) \leq c_2$, we obtain $|\lambda'_{k+1}| \leq c_2|\lambda^2_{k+1} \beta'_{k+1}|$.

To complete the induction, we establish (4). Consider the one-step map $\beta_{k+1}(\beta_k^{-1}(\,\cdot \,))$. On $\calB(\frac{1}{3})$, this has been shown to be defined with bounds on $\beta'_{k+1}$. We have already estimated $\beta'_k$ on $\beta^{-1}_{k-1}(\calB(\frac{1}{3}))$, which is larger than $\beta^{-1}_k(\calB(\frac{1}{3}))$, by (4). Hence the composition has derivative $L^2+O(\lambda_{k-1})$. In  addition, the recursion (\ref{equation1.16}) shows that $\beta_{k+1}(\beta^{-1}_k(0))$ is $O(\lambda_k)$. Hence $\beta_{k+1}(\beta_k^{-1}(\calB(\frac{1}{3})))$ covers $\calB(\frac{1}{2})$ and so $\beta^{-1}_{k+1}(\calB(\frac{1}{2})) \subset \beta^{-1}_k(\calB(\frac{1}{3}))$. Chaining this inclusion down to $k=0$, we obtain (4), and the proof is complete.
\hspace*{20pt} $\Box$
\bigskip

{\em Proof of Proposition \ref{proposition1.3}}.
We may define
$$
\beta^c(\lambda) = \dis\bigcap^\infty_{k=0} \beta_k^{-1}(\calB(\frac{1}{2})),
$$
since Lemma \ref{lemma4.1}(3, 4) imply that these sets are a decreasing sequence of open sets with diameter $\leq cL^{-2k} l_k^{\frac{1}{4}}$. Furthermore, at $\beta^c(\lambda)$, Lemma 4.1(2) holds for all $k$, so $\lambda_k(\beta^c(\lambda)) \rarrow 0$ as $k \rarrow \infty$.
Consider the sequence $\beta^c_n = \beta_n(\beta^c(\lambda))$. By construction, this is a bounded sequence obeying $\beta^c_{n+1} = L^2\beta^c_n+O(\lambda_n)$ ({\em c.f.}~(\ref{equation1.16})) and as such it must satisfy $\beta^c_n = O(\lambda_n) \rarrow 0$. In particular, $\beta^c(\lambda) = \beta^c_0 = O(\lambda)$.

In order to complete the proof of Proposition \ref{proposition1.3}, we compute the shifted recursion which applies to $\hat{\beta} = \beta -\beta^c(\lambda)$. Let $\hat{\beta}_j(\hat{\beta}) = \beta_j(\hat{\beta}+\beta^c(\lambda)) - \beta^c_j$ denote the difference between the flow from $\beta$ and the critical flow from $\beta^c(\lambda)$. Then (\ref{equation1.16}) becomes
\begin{eqnarray*}
\lambda_{j+1} & = & \lambda_j - \, \frac{8B\lambda^2_j}{(1+\hat{\beta}_j+\beta^c_j)^2}
 + \epsilon_{\lambda,j}, \\[4mm]
\hat{\beta}_{j+1} & = & L^2\left[\hat{\beta}_j+2B\left(\frac{1}{1+\hat{\beta}_j+\beta^c_j} - \frac{1}{1+\beta^c_j}\right)\lambda_j\right] + \epsilon_{\beta,j}(\hat{\beta}+\beta^c(\lambda)) - \epsilon_{\beta,j}(\beta^c(\lambda)).
\end{eqnarray*}

We control the global behavior of this recursion with another lemma. Some additional definitions will be needed. Let $k_{\hat{\beta}}$ be the largest $k$ such that $|\hat{\beta}_k| \leq 1$ (if no such integer exists, then $k_{\hat{\beta}} =0$). Then with $\hat{k} = \min\{k,k_{\hat{\beta}}\}$, we define
\begin{equation}\label{equation4.11}
l_k(\hat{\beta}) = \exp\left[\dis\sum^{\hat{k}-1}_{j=0} \, \frac{8B}{\lambda^{-1}+8Bj}\right],
\end{equation}
and observe that $|l_k(\hat{\beta})| = l_{\hat{k}}$, {\em c.f.}~(\ref{equation4.1}).  Again, integral approximation shows that $l_k(\hat{\beta}) = (1+8B\lambda \hat{k})e^{O(\lambda)}$.

\begin{lemma}\label{lemma4.2}
Let $\calD_\beta(\rho) = \calD_\beta + \calB(\rho)$. Then for $(\hat{\beta},\lambda) \in \calD_\beta(\frac{1}{4} L^{-2k} l_k^{\frac{1}{4}})\times \calD_\lambda$, the following bounds hold with $k$-independent constants:
\begin{itemize}
\item[(1)]
$\lambda_k \in \overline{\calD}_\lambda$ and
$$
\left|\lambda_k - \, \frac{1}{\lambda^{-1} +8B\hat{k}}\right| \leq c_1 \left|\frac{1}{\lambda^{-1}+8B\hat{k}}\right|^2(1+\ln(1+|\lambda|\hat{k})).
$$
\item[(2)]
$\hat{\beta}_k = \hat{\beta} L^{2k} l_k(\hat{\beta})^{-\frac{1}{4}} e^{O(\lambda)} \in \overline{\calD}_\beta(\frac{1}{3})$. If $\hat{\beta} \in \calD_\beta$, then $\hat{\beta}_k \in \overline{\calD}_\beta$.
\item[(3)]
$|\lambda'_k| \leq c_2|\lambda^2_k \beta'_k| \ |1+\hat{\beta}_k|^{-1}, \ \beta'_k = L^{2k} l_k(\hat{\beta})^{-\frac{1}{4}} e^{O(\lambda)}$.  (Note that $\hat{\beta}'_k=\beta'_k$.)
\item[(4)]
The recursion relations
\begin{eqnarray*}
\hat{\beta}_{k+1} & = & L^2 \hat{\beta}_k \left(1 - \, \frac{2B \lambda_k}{1+\hat{\beta}_k} + \hat{\epsilon}_{\beta,k}\right), \\[4mm]
\lambda_{k+1} & = & \lambda_k - \, \frac{8B \lambda^2_k}{(1+\hat{\beta}_k)^2} + \hat{\epsilon}_{\lambda,k},
\end{eqnarray*}
hold with $\hat{\epsilon}_{\beta,k}, \hat{\epsilon}_{\lambda,k}$ analytic in $\hat{\beta}$ and satisfying
$$
|\hat{\epsilon}_{\beta,k}| \leq c_3|\lambda_k|^2 \ |1+\hat{\beta}_k|^{-1}, \ |\hat{\epsilon}_{\lambda,k}| \leq c_4|\lambda_k|^3  |1+\hat{\beta}_k|^{-1}.
$$
In addition, for $\hat{\beta} \in \calD_\beta \left(\frac{1}{5} L^{-2k} l_k^{\frac{1}{4}}\right)$,
$$
|\hat{\epsilon}'_{\beta,k}| \leq c_5|\lambda^2_k\beta'_k| \
|1+\hat{\beta}_k|^{-2}, \ 
|\hat{\epsilon}'_{\lambda,k}| \leq c_6|\lambda^3_k\beta'_k| \  |1+\hat{\beta}_k|^{-1}. 
$$
\end{itemize}
\end{lemma}

Lemma \ref{lemma4.2} shows that if $\beta \in \calD_\beta + \beta^c(\lambda)$ and $\lambda \in \calD_\lambda$, then (2) holds for all $k$. Thus $\beta_k \in \overline{\calD}_\beta + \beta^c_k$, which completes the proof of Proposition \ref{proposition1.3}.  \hspace*{20pt} $\Box$
\bigskip

{\em Proof of Lemma \ref{lemma4.2}}. We begin by showing (1), (2), (3) imply (4). We may assume Lemma \ref{lemma4.2} for smaller values of $k$. Since $(\hat{\beta}_j,\lambda_j) \in \overline{\calD}_\beta(\frac{1}{3}) \times \overline{\calD}_\lambda$ for $j = 1,\ldots,k$, and since $\hat{\beta}_j - \beta_j = \beta^c_j = O(\lambda)$, the assumption in Proposition II.6.1 holds and the recursion relations (\ref{equation1.16}), (\ref{equation1.17}) are valid.

As $\beta^c_k = O(\lambda_k)$, and as $1+\hat{\beta}_k$ is never going near 0, we can expand in $\beta^c_k$ in the $\lambda$ recursion, with all but the zero$^{\rm{th}}$ order going into the remainder. For the $\beta$ recursion, we write
$$
\frac{1}{1+\hat{\beta}_k+\beta^c_k} \, - \, \frac{1}{1+\beta^c_k} =
\frac{-\hat{\beta}_k}{(1+\beta^c_k)(1+\hat{\beta}_k+\beta^c_k)} =
\frac{-\hat{\beta}_k}{1+\hat{\beta}_k} 
+ \frac{\hat{\beta}_k \beta^c_k(2+\hat{\beta}_k+\beta^c_k)}
{(1+\hat{\beta}_k)(1+\beta^c_k)(1+\hat{\beta}_k+\beta^c_k)},
$$
with the second term going into the remainder, as it is proportional to $\beta^c_k = O(\lambda_k)$. The result is
\begin{eqnarray*}
\lambda_{k+1} & = & \lambda_k - \, \frac{8B \lambda_k^2}{(1+\hat{\beta}_k)^2} + \hat{\epsilon}_{\lambda,k} \nonumber \\[4mm]
\hat{\beta}_{k+1} & = & L^2 \hat{\beta}_k\left[1- \, \frac{2B \lambda_k}{1+\hat{\beta}_k} + \hat{\epsilon}_{\beta,k}\right],
\end{eqnarray*}
with $\hat{\epsilon}_{\lambda,k}$ still of order $|\lambda_k|^3|1+\beta_k|^{-1} \approx |\lambda_k|^3|1+ 
\hat{\beta}_k|^{-1}$, and with
\begin{equation}\label{equation4.12}
\hat{\epsilon}_{\beta,k} = 2B \lambda_k 
\beta^c_k \, \frac{2+\hat{\beta}_k +\hat{\beta}^c_k}
{(1+\hat{\beta}_k)(1+\beta^c_k)(1+\hat{\beta}_k+\beta^c_k)}
 + 
\frac{\epsilon_{\beta,k}(\hat{\beta}+\beta^c(\lambda))-\epsilon_{\beta,k}(\beta^c(\lambda))}{\hat{\beta}_k}.
\end{equation}

The first term in $\hat{\epsilon}_{\beta,k}$ is $O(\lambda^2_k)|1+\hat{\beta}_k|^{-1}$. To bound the second term, consider two cases. First, if $|\hat{\beta}_k| < \frac{1}{10}$, then write the second term as
$$
\int_0^1 d\theta \frac{\hat{\beta}}{\hat{\beta}_k} \, \epsilon'_{\beta,k}(\theta \hat{\beta}+\beta^c(\lambda)).
$$
Note that in this case $\hat{k} = k$, $l_k = |l_k(\hat{\beta})|$, so (2) implies that $\hat{\beta} \in \calB \left(\frac{1}{10} L^{-2k} l_k^{\frac{1}{4}} e^{O(\lambda)}\right)$.
Double the size of this ball, so that Cauchy's bound may be used. 
To check the assumptions of Proposition II.6.1, observe that for $0 \le j \le k$, $\calB\left(\frac{1}{5} L^{-2k} l_k^{\frac{1}{4}} e^{O(\lambda)}\right)
\subset \calD_\beta\left(\frac{1}{4} L^{-2j}l_j^{\frac{1}{4}}\right)$, so that (2) holds, and in particular $\hat{\beta}_j \in \calB(\frac{1}{3})$.  Hence  (\ref{equation1.17}) holds and $|\epsilon_{\beta,k}| \leq O(\lambda^2_k)$. Cauchy's estimate then implies
$$
\left|\frac{\hat{\beta}}{\hat{\beta}_k} \, \epsilon'_{\beta,k} (\theta \hat{\beta} + \beta^c(\lambda))\right|
\leq \left|\frac{\hat{\beta}}{\hat{\beta}_k}\right| O(\lambda^2_k)L^{2k} l_k^{-\frac{1}{4}} \leq O(\lambda^2_k).
$$
In the second case $(|\hat{\beta}_k| \geq \frac{1}{10})$ each $\epsilon_{\beta,j}$ term can be estimated separately.  
Note that Lemma \ref{lemma4.2}(2) applies for $0 \leq j \leq k$, since $\calD_\beta(\frac{1}{4} L^{-2k}l_k^{\frac{1}{4}})$ is decreasing in $k$.
Hence (\ref{equation1.17}) holds, so that
\begin{equation}
\label{equation4.13}
\left| \frac{\epsilon_{\beta,k} (\hat{\beta}+\beta^c(\lambda))-\epsilon_{\beta,k}(\beta^c(\lambda))}{\hat{\beta}_k}\right|
\leq O(\lambda^2_k)|1+\hat{\beta}_k|^{-2}.
\end{equation}

Proceeding to the derivatives, we use Cauchy's estimate with the bounds just established on $\hat{\epsilon}_{\beta,k}$, $\hat{\epsilon}_{\lambda,k}$. Thus if we shrink the domain to $\calD_\beta(\frac{1}{5} L^{-2k} l_k^{\frac{1}{4}})$, we have
\begin{eqnarray*}
|\hat{\epsilon}'_{\lambda,k}| & \leq &
c|\lambda_k|^3|1+\hat{\beta}_k|^{-1} L^{2k} l_k^{-\frac{1}{4}} \leq c|\lambda_k|^3|1+\hat{\beta}_k|^{-1}|\beta'_k|, \\[4mm]
|\hat{\epsilon}'_{\beta,k}| & \leq &
c|\lambda_k|^2|1+\hat{\beta}_k|^{-2} L^{2k} l_k^{-\frac{1}{4}} \leq c|\lambda_k|^2|1+\hat{\beta}_k|^{-2}|\beta'_k|, \end{eqnarray*}
where we have used (3) and $|l_k(\hat{\beta})/l_k| \leq O(1)$ to relate $|\beta'_k|$ to $L^{2k}l_k^{-\frac{1}{4}}$. The $\hat{\epsilon}'_{\beta,k}$ bound was obtained by differentiating the first term in (\ref{equation4.12})
explicitly, and using (\ref{equation4.13}) on the second term.  This completes the proof of (4). It also gets the induction started, since (1), (2), (3) are trivial for $k=0$.

To complete the cycle, we show that (4) (with $k+1$ replaced by $k$) implies (1), (2), and (3). To prove (1), proceed as in (\ref{equation4.2})-(\ref{equation4.3}). In this case we have
$$
\sum^{k-1}_{j=0} \left[
\frac{8B}{(1+\hat{\beta}_j)^2} 
+ \frac{O(\lambda_j)}{|1+\hat{\beta}_j|}\right] = 8B \hat{k}+O(1)(1+\ln(1+|\lambda|\hat{k})),
$$
and the bound in (1) follows. The argument for $\lambda_k \in \overline{\calD}_\lambda$ is unchanged. To obtain (2), express the iteration of (4) in exponential form:
$$
\hat{\beta_k} = \hat{\beta_k} L^{2k} \exp\left[\sum^{k-1}_{j=0} \left(\frac{-2B\lambda_j}{1+\hat{\beta}_j} + \frac{O(\lambda^2_j)}{1+\hat{\beta}_j}\right)\right].
$$
The geometric growth of $\hat{\beta}_j$ and (1) show that this may be expressed as in (2).

In order to prove that $\hat{\beta}_k \in \overline{\calD}_\beta(\frac{1}{3})$, we need to allow for the phase change from $l_k(\hat{\beta})^{-\frac{1}{4}}$ in the bound of (2). Since $l_k(\hat{\beta}) = (1+8B \lambda \hat{k})e^{O(\lambda)}$, we have $|\mbox{arg } l_k(\hat{\beta})| \leq |\mbox{arg }\lambda|+O(\lambda)$. Thus if $|\mbox{arg } \hat{\beta}| < b_\beta$, then $|\mbox{arg } \hat{\beta}_k| < b_\beta + \frac{1}{4} \, b_\lambda +O(\lambda)$, so that $\hat{\beta}_k \in \overline{\calD}_\beta$ for all $\hat{\beta} \in \calD_\beta$.

Before we may conclude that $\hat{\beta}_k \in \overline{\calD}_\beta(\frac{1}{3})$ for all $\hat{\beta} \in \calD_\beta(\frac{1}{4} L^{-2k}l_k^\frac{1}{4})$, we need to allow for the spilling out of $\hat{\beta}_k$ from $\calD_\beta(\frac{1}{4})$ due to the slow variation of $l_k(\hat{\beta})$ with $\hat{\beta}$ in the bound of (2).  Consider a ball of radius $\frac{1}{4} L^{-2k}l_k^{\frac{1}{4}}$ and centered at $\hat{\beta} \in \calD_\beta$.  The bound in (2) shows that in the $\hat{\beta}_k$ plane, it scales up to an approximate ball of radius $\frac{1}{4}|l_k^{\frac{1}{4}}/l_k(\hat{\beta})^{\frac{1}{4}}|=\frac{1}{4}|l_k^{\frac{1}{4}}/l_{\hat{k}}^{\frac{1}{4}}|$.  As this ball may be larger than the ball of radius $\frac{1}{4}$ centered at $\hat{\beta}_k$, some widening of the opening angle in $\calD_\beta(\frac{1}{3})$ is needed.  This is only a problem if $k > \hat{k} \equiv \min \{k,k_{\hat{\beta}}\}$, in which case $\hat{\beta}_k > 1$, by the definition of $k_{\hat{\beta}}$.  We claim that $|l_k^{\frac{1}{4}}/l_{\hat{k}}^{\frac{1}{4}}|-1 \le O(\lambda)|\hat{\beta}_k|$, which implies that an $O(\lambda)$ increase in opening angle is sufficient.  For a proof, observe first that
$|\hat{\beta}_k| \geq c^{-1} L^{k-k_{\hat{\beta}}}$.  This is a consequence of the fact that $\hat{\beta}_k$ has geometric growth with ratio close to $L^2$, and the fact that by definition, $\hat{\beta}_{k_{\hat{\beta}}}$ is no smaller than $L^{-2}(1+O(\lambda)) = c^{-1}$.  Second, a crude estimate on (\ref{equation4.1}) gives 
$$
|l_k^{\frac{1}{4}}/l_{\hat{k}}^{\frac{1}{4}}| \leq e^{O(\lambda)(k-\hat{k})} \leq 
|c \hat{\beta}_k|^{O(\lambda)}.
$$
Letting $y=\ln |c\hat{\beta}_k|$, we may use the fact that $e^{ay}-1 < ae^y$ for $a, y \ge 0$ to conclude that
$|l_k^{\frac{1}{4}}/l_{\hat{k}}^{\frac{1}{4}}|-1 \le  O(\lambda)|\hat{\beta}_k|$ as claimed.  As a result, we have that $|\hat{\beta}_k-z|<\frac{1}{3}$ for some $z$ with $|\mbox{arg } z| < b_\beta + \frac{1}{4} \, b_\lambda +\epsilon$, and so $\hat{\beta}_k \in \overline{\calD}_\beta(\frac{1}{3})$.

We proceed to the proof of (3). Differentiating (4), we obtain
\begin{eqnarray*}
\lambda'_{k+1} & = & \lambda'_k - \, \frac{16B(\lambda_k\lambda'_k-\lambda^2_k(1+\hat{\beta}_k)^{-1} \beta'_k)}{(1+\hat{\beta}_k)^2} + \hat{\epsilon}'_{\lambda,k}, \\[4mm]
\beta'_{k+1} & = & L^2 \beta'_k \left[1- \, \frac{2B}{1+\hat{\beta}_k}\left(\lambda_k + \frac{\lambda'_k \hat{\beta}_k}{\beta'_k} - \frac{\hat{\beta}_k\lambda_k}{1+\hat{\beta}_k}\right) + \hat{\epsilon}_{\beta,k} + \frac{\hat{\epsilon}'_{\beta,k} \hat{\beta}_k}{\beta'_k}\right].
\end{eqnarray*}
From (3) (applied to $\lambda'_k$) and (4) we see that
$$
|\lambda'_{k+1}|  = \frac{|\lambda^2_k \beta'_k|}{|1+\hat{\beta}_k|}
(c_2+O(\lambda_k)+O(1)),
$$
and as before, {\em c.f.}~(\ref{equation4.10}), by choosing $c_2$ large enough we obtain the desired bound on $\lambda'_{k+1}$. Likewise we apply the inductive assumptions to each term in the $\beta'_{k+1}$ equation to obtain
$$
\beta'_{k+1} = L^2 \beta'_k\left[1- \, \frac{2B \lambda_k}{1+\hat{\beta}_k} + \frac{O(\lambda^2_k)\hat{\beta}_k}{|1+\hat{\beta}_k|^2 }
+ \frac{O(\lambda_k)\hat{\beta}_k}{|1+\hat{\beta}_k|^{2}} +
\frac{O(\lambda^2_k)}{|1+\hat{\beta}_k|}\right].
$$
We put this in exponential form:
\begin{eqnarray*}
\beta'_{k+1} & = & L^{2(k+1)} \exp\left[\sum^{\hat{k}}_{j=0} \left(-2B \lambda_k+ \frac{O(\hat{\beta}_k\lambda_k)}{|1+\hat{\beta}_k|}\right)\right]e^{O(\lambda)} \\[4mm]
& = & L^{2(k+1)} l_k(\hat{\beta})^{-\frac{1}{4}}e^{O(\lambda)}.
\end{eqnarray*}
The error from replacing $\lambda_k$ with $(\lambda^{-1}+8B\hat{k})^{-1}$ (as with all the other error terms) is summable to $O(\lambda)$. \hspace*{20pt} $\Box$

\begin{corollary}\label{corollary4.3}
If $(\hat{\beta},\lambda) \in \calD_\beta \times \calD_\lambda$, then
\begin{equation}\label{equation4.14}
k_{\hat{\beta}} = O(1) + \frac{1}{2} \log(1+|\hat{\beta}|^{-1}) + \frac{1}{8} \log|1+4B \lambda \log(1+|\hat{\beta}|^{-1})|.
\end{equation}
\end{corollary}

{\em Proof.} If $|\hat{\beta}| \geq 1$, then $k_{\hat{\beta}}=0$ and (\ref{equation4.14}) is valid. If $|\hat{\beta}| <1$, then 
we need to solve for $k$ in the equation $\hat{\beta}_k=O(1)$.  By Lemma \ref{lemma4.2}(2) and the fact that 
$|l_k(\hat{\beta})| = |1+8B \lambda \hat{k}|e^{O(\lambda)}$, this can be written as
$$
|\hat{\beta}_k|L^{2k}|1+8B\lambda k|^{-\frac{1}{4}}=O(1).
$$
Rewrite this as
$$
k = O(1) + \frac{1}{2} \log|\hat{\beta}|^{-1} + \frac{1}{8} \log|1+8B \lambda k|,
$$
and solve by repeated substitution.  The result can be expressed as in 
(\ref{equation4.14}).

\hspace*{20pt} $\Box$ 
\bigskip

{\em Proof of Proposition 1.5.} (1) is just the shifted version of a statement in Proposition \ref{proposition1.3}. (2) is Corollary \ref{corollary4.3}. (3) is a restatement of Lemma \ref{lemma4.2}(1). To obtain (4), note that by Lemma \ref{lemma4.2}(2),
\begin{equation}\label{equation4.15}
\ell_k(\beta)^{-\frac{1}{4}} = \beta_k L^{-2k}/ \hat{\beta} = l_k(\hat{\beta})^{-\frac{1}{4}} e^{O(\lambda)} = (1+8B \lambda \hat{k})^{-\frac{1}{4}} e^{O(\lambda)}.
\end{equation}
(5) follows immediately from the geometric growth of $\hat{\beta}_k$ and the recursion relation and bounds in Lemma \ref{lemma4.2}(4). To obtain (6), consider first the ratio
\begin{equation}\label{equation4.16}
\left|\frac{\ell_k(T^{-1})}{\ell_k(\hat{\beta})}\right| = e^{O(\lambda)} \left|\frac{1+8B \lambda \hat{k}_1}{1+8B \lambda \hat{k}_2}\right|,
\end{equation}
where $\hat{k}_1 = \min \{k,k_{1/T}\}$ and $\hat{k}_2 = \min\{k,k_{\hat{\beta}}\}$. By Corollary \ref{corollary4.3},  if $|\hat{\beta} T| > 1$, then
$$
O(1) \leq k_{1/T}-k_{\hat{\beta}} \leq O(1) + \left(\frac{1}{2} + \epsilon\right) \log |\hat{\beta}T|.
$$
The same bounds hold for $\hat{k}_1-\hat{k}_2$, so (\ref{equation4.16}) implies
$$
e^{O(\lambda)} \leq \left|\frac{\ell_k(T^{-1})}{\ell_k(\hat{\beta})}\right| \leq e^{O(\lambda)}(1+O(\lambda)(1+\log |\hat{\beta}T|)).
$$
To get the same bounds on $|\lambda_k(\hat{\beta})/\lambda_k(T^{-1})|$, note that Lemma \ref{lemma4.2}(1) and (\ref{equation4.15}) imply
$$
\lambda_k{(\hat{\beta})} = \frac{\lambda}{1+8B \lambda \hat{k}} = \frac{\lambda}{\ell_k(\hat{\beta})} \ e^{O(\lambda)},
$$
so the $\lambda_k(\hat{\beta})/\lambda_k(T^{-1})$ bound is really the same as the $\ell_k(T^{-1})/\ell_k(\hat{\beta})$ bound.

To obtain (7), apply the recursion relations of Lemma \ref{lemma4.2}(4) 
{\em ad infinitum}:
\begin{eqnarray*}
\ell(\hat{\beta})
& = & \dis\prod^\infty_{k=0} \left(1 - \, \frac{2B \lambda_k}{1+\hat{\beta}_k} + \hat{\epsilon}_{\beta,k}\right) \\[4mm]
(\ln \ell(\hat{\beta}))' & = & \sum^\infty_{k=0} \left[\frac{-2B \lambda'_k}{1+\hat{\beta}_k}
+ \frac{2B \lambda_k \beta'_k}{(1+\hat{\beta}_k)^2} + \hat{\epsilon}'_{\beta,k}\right] e^{O(\lambda)}
 =  \sum^\infty_{k=0} \frac{O(\lambda_k)\beta'_k}{(1+\hat{\beta}_k)^2}  .
\end{eqnarray*}
By Lemma \ref{lemma4.2}(2,3), we have $\beta'_k = \hat{\beta}_k \hat{\beta}^{-1}e^{O(\lambda)}$, so this can be written as
$$
\hat{\beta}^{-1} \sum^\infty_{k=0} O(\lambda_k) \hat{\beta}_k|1+\hat{\beta}_k|^{-2} = \hat{\beta}^{-1} O(\lambda_{k_{\hat{\beta}}}) = \hat{\beta}^{-1}O(\lambda_{\hat{\beta}}),
$$
and (7) is an immediate consequence.

Proceeding to (8), note that Lemma \ref{lemma4.2}(4) implies that
$$
\frac{\beta_{{\rm eff},\infty}}{\beta_{{\rm eff},k}} = \prod^\infty_{j=k} \left(1- \, \frac{2B \lambda_j}{1+\hat{\beta}_j} + \hat{\epsilon}_{\beta,j}\right).
$$
Thus we may obtain (8) from the following sequence of bounds:
\begin{eqnarray*}
\left|\ln \, \frac{\beta_{{\rm eff},k}}{\beta_{{\rm eff},\infty}}\right|
& = & \sum^\infty_{j=k} \left[\frac{2B \lambda_j}{1+\hat{\beta}_j} + \frac{O(\lambda^2_j)}{1+\hat{\beta}_j}\right] \\[4mm]
& \leq & \sum^{k_{\hat{\beta}}}_{j=k} O(\lambda_j) + \sum^\infty_{j=k_{\hat{\beta}}} \frac{O(\lambda_{\hat{\beta}})}{|1+\hat{\beta}_j|} \\[4mm]
& \leq & O(\lambda_k)(1+\max\{k_{\hat{\beta}}-k,0\}) \\[4mm]
& \leq & O(\lambda_k) (1+ \log(1+ |\hat{\beta}_k|^{-1} )).
\end{eqnarray*}
In the last step we have used the fact that since $k_{ \hat{\beta} }$ is defined so that $\hat{\beta}_{k_{\hat{\beta}}}\leq 1$, the recursion implies that $|\hat{\beta}_k| \leq L^{-(2-\epsilon)(k_{\hat{\beta}}-k)}$ for $k < k_{\hat{\beta}}$. 
$\hspace*{20pt} \Box$


\begin{thebibliography}{BLZ73}

\bibitem[BEI92]{BEI}
Brydges, D. C., Evans, S.~E., and Imbrie, J.~Z.: Self-avoiding walk on a hierarchical lattice in four dimensions. {\em Ann.~Probab.}~{\bf 20}, 82--124 (1992).

\bibitem[BI02]{BI2}
Brydges, D. C., and Imbrie, J.~Z.: The Green's function for a hierarchical self-avoiding walk in four dimensions.  Preprint, arXiv:math-ph/0205028.

\bibitem[BLZ73]{BLZ}
Br\'ezin, E., Le Guillou, J.~C., and Zinn-Justin, J.: Approach to scaling in renormalized perturbation theory. {\em Phys.~Rev.}~{\bf D8}, 2418--2430 (1973).

\bibitem[F71]{F}
Feller, W.: {\em An Introduction to Probability Theory and its Applications, Vol.~II}, 2nd ed. New York: Wiley (1971).

\bibitem[G94]{G}
Golowich, S. E.: Rigorous results for self-avoiding random walks.  Harvard Ph.D. thesis, 1994.

\bibitem[GI95]{GI}
Golowich, S. E., Imbrie, J. Z.:  The broken supersymmetry phase of a self-avoiding walk.  {\em Commun. Math. Phys.} {\bf 168}, 265-320 (1995).

\bibitem[HS92]{HS}
Hara, T., Slade, G.: Self-avoiding walk in five or more dimensions. I. The critical behavior. {\em Commun.~Math.~Phys.}~{\bf 147}, 101--136 (1992). The lace expansion for self-avoiding walk in five or more dimensions. {\em Rev.~Math.~Phys.}~{\bf 4}, 235--327 (1992).

\bibitem[HT02]{HT}
Hattori, T., Tsuda, T.:  Renormalization group analysis of the self-avoiding paths on the $d$-dimensional Sierpi\'{n}ski gasket.  Preprint, mp\_arc:02-255.          

\bibitem[IM94]{IM}
Iagolnitzer, D., and Magnen, J.: Polymers in a weak random potential in dimension four: Rigorous renormalization group analysis. {\em Commun.~Math.~Phys.}~{\bf 162}, 85--121 (1994).

\bibitem[MS93]{MS}
Madras, N., Slade, G.: {\em The self-avoiding walk}. Boston: Birkh\"auser, 1993.

\end{thebibliography}
\end{document}